\documentclass{article}
\usepackage{amssymb,stmaryrd}
\usepackage{amsmath,bm,shuffle}
\usepackage{array}
\usepackage{caption,graphicx}
\usepackage[all]{xy}
\usepackage{color}

\newcommand{\be}{\begin{equation}}
\newcommand{\ee}{\end{equation}}

\def\un{{\rm 1\mkern-4mu I}}

\title{Octonion Internal Space Algebra for the Standard Model\footnote{Extended version of a lecture presented at the Workshop \textit{Octonions and the Standard Model}, Perimeter Institute, Waterloo, Canada, February-May 2021, and at the 14th International Workshop \textit{Lie Theory and Its Applications to Physics} (LT 14), Sofia, June 2021.}}
\author{Ivan Todorov}
\date{
\small Institut des Hautes \'Etudes Scientifiques, 35 route de Chartres, 
\\ F-91440 Bures-sur-Yvette, France
\\ and
\\ Institute for Nuclear Research and Nuclear Energy, Bulgarian Academy of Sciences, 
\\ Tsarigradsko Chaussee 72, BG-1784 Sofia, Bulgaria
\\ (permanent address)
\\e-mail address: ivbortodorov@gmail.com
}

\begin{document}

\maketitle

\vglue 10mm

\begin{abstract}
The paper surveys recent progress in the search for an appropriate internal space algebra for the Standard Model (SM) of particle physics. As a starting point serve Clifford algebras involving operators of left multiplication by octonions. A central role is played by a distinguished complex structure which implements the splitting of the octonions ${\mathbb O} = {\mathbb C} \oplus {\mathbb C}^3$ reflecting the lepton-quark symmetry. Such a complex structure in  $C\ell_{10}$ is generated by the $C\ell_6(\subset C\ell_8\subset C\ell_{10})$ volume form, $\omega_6 = \gamma_1 \cdots \gamma_6$, left invariant by the Pati-Salam subgroup of $Spin(10)$, $G_{\rm PS} = Spin (4) \times Spin (6) / {\mathbb Z}_2$. While the $Spin(10)$ invariant volume form $\omega_{10}=\gamma_1 ... \gamma_{10}$ is known to split the Dirac spinors of $C\ell_{10}$  into left and right chiral (semi)spinors,  ${\cal P} = \frac12 (1 - i\omega_6)$ is interpreted as the projector on the 16-dimensional \textit{particle subspace} (annihilating the antiparticles).  The standard model gauge group appears as the subgroup of $G_{PS}$ that preserves the sterile neutrino (identified with the Fock vacuum). The $\mathbb{Z}_2$-graded internal space algebra $\mathcal{A}$ is then included in the projected  tensor product: $\mathcal{A}\subset \mathcal{P}C\ell_{10}\mathcal{P}=C\ell_4\otimes \mathcal{P} C\ell_6^0\mathcal{P}$.  The Higgs field appears as the scalar term of a superconnection, an element of the odd part, $C\ell_4^1$, of the first factor. The fact that the projection of $C\ell_{10}$ only involves the even part $C\ell_6^0$ of the second factor guarantees that the colour symmetry remains unbroken. As an application we express the ratio $\frac{m_H}{m_W}$ of the Higgs to the $W$-boson masses in terms of the cosine of the {\it theoretical} Weinberg angle.
\end{abstract}
Key words: Clifford algebra, complex structure, superconnection, Higgs mass 
\vfill\eject

\tableofcontents

\newpage

\section{Introduction}\label{sec1}

The elaboration of the Standard Model (SM) of particle physics was completed in the early 1970's. To quote John Baez \cite{B21} 50 ``years trying to go beyond the Standard Model hasn't yet led to any clear success''. The present survey belongs to an equally long albeit less fashionable effort - to clarify the algebraic (or geometric) roots of the SM, more specifically, to find a natural framework featuring its internal space properties. After discussing some old explorations we provide an updated exposition of recent developments (in particular of \cite{T21}), clarifying on the way the meaning and the role of complex structures, concentrating on one associated with a Clifford subalgebra (in our case $C\ell_6$) pseudoscalar. 

\smallskip

Most ideas on the natural framework of the SM originate in the 1970's, the first decade of its existence. (Two exceptions: the Jordan algebras were introduced and classified in the 1930's \cite{J,JvNW}; the noncommutative geometry approach was born in the late 1980's, \cite{C,DKM,CL}, and is still vigorously developed by Connes, collaborators and followers \cite{CC,CCS,BF, CIS,NS}.)

\smallskip

First, early in 1973, the ultimate division algebra, the octonions\footnote{For a pleasant to read review of octonions, their history and applications -- see \cite{B02}.} were introduced by G\"ursey\footnote{I had the good fortune to know him personally. See Witten's eloquent characterization of his personality and work in the Wikipedia entry on Feza G\"ursey (1921-1991).} and his student G\"unaydin \cite{GG,G} for the description of quarks and their $SU(3)$ colour symmetry. The idea was taken up and extended to incorporate all four division algebras by G. Dixon (see \cite{D10,D14} and earlier work cited there) and is further developed by Furey \cite{F14,F15,F16,F,F18,FH1,FH}. Dubois-Violette (D-V) \cite{D16} emphasizes the lepton-quark correspondence and the unimodularity of the colour group, $SU(3)_c$, as a physical motivation for introducing the octonions. They come equipped with a complex structure preserved by the subgroup $SU(3)$ of the automorphism group $G_2$ of ${\mathbb O}$:
\be
\label{eq11}
{\mathbb O} = {\mathbb C} \oplus {\mathbb C}^3 \, .
\ee

\subsection{(Split) octonions as composition algebras}

One can in fact provide a basis free definition of the octonions starting with the splitting (\ref{eq11}). To this end one uses the skew symmetric vector product and the standard inner product on ${\mathbb C}^3$ to define a noncommutative and non-associative distributive product $xy$ on ${\mathbb O}$ and a real valued nondegenerate symmetric bilinear form $\langle x,y \rangle = \langle y,x \rangle$ such that the quadratic norm $N(x) = \langle x,x \rangle$ is multiplicative:  
\be
\label{eq12}
N(xy) = N(x) N(y) \quad \mbox{for} \quad N(x) = \langle x,x \rangle \, 
\ee
(cf. \cite{D16,TD}). Furthermore, defining the real part of $x \in {\mathbb O}$ by ${\rm Re} \, x = \langle x,1 \rangle$ and the octonionic conjugation $x \to x^* = 2 \langle x,1 \rangle - x$, we shall have
\be
\label{eq13}
xx^* = N(x)1 \Leftrightarrow x^2 - 2 \langle x,1 \rangle x + N(x)1 = 0,\,
\ee
where $1$ stands for the (real) octonion unit. A unital algebra with a non-degenerate quadratic norm obeying (\ref{eq12}) is called a {\it composition algebra}.

\smallskip

{\footnotesize Another basis free definition of the octonions ${\mathbb O}$ and of their split version $\widetilde{\mathbb O}$ can be given in terms of quaternions by the Cayley-Dickson construction. We represent the quaternions as scalars plus vectors
$$
{\mathbb H} = {\mathbb R} \oplus {\mathbb R}^3 , \ x = u + U , \ y = v + V , \ u,v \in {\mathbb R} , \ U,V \in {\mathbb R}^3 ,
$$
\be
\label{eq14}
xy = uv - \langle U,V \rangle + uV + U v + U \times V
\ee
with the vector product $U \times V \in {\mathbb R}^3$ satisfying
\be
\label{eq15}
U \times V = - V \times U , \ (U \times V) \times W = \langle U,W \rangle V - \langle V,W \rangle U \, .
\ee
The product (\ref{eq14}) is clearly noncommutative but one verifies that it is associative. The Cayley-Dickson construction defines the octonions ${\mathbb O}$ and the split octonions $\widetilde{\mathbb O}$ in terms of a pair of quaternions and a new  ``imaginary unit'' $l$ as:
$$
x = u + U + l (v + V) , \ l (v+V) = (v - V) l \, ,
$$
\be
\label{eq16}
l^2 = \left\{\begin{matrix}
-1 &\Rightarrow &x \in {\mathbb O}\hfill \\
\hfill 1 &\Rightarrow & \ x \in \widetilde{\mathbb O} \, . 
\end{matrix} \right.
\ee 
We shall encounter the split octonions as generators of $C\ell(4, 2)$ in Sect. 3.1 below.}

\subsection{Jordan algebras; GUTs; Clifford algebras}

Studying quantum field theory it appears natural to replace  classical observables (real valued functions) by an algebra of functions on space-time with values in a {\it finite dimensional euclidean Jordan algebra}\footnote{These algebras are defined and classified in \cite{J, JvNW}; for a concise review see \cite{M11} as well as Sect.~3.2 in \cite{D16} and Sect.~2 of \cite{T}; about Pascual Jordan (1902-1980) see \cite{Da}.}. As a particularly attractive choice, which incorporates the idea of lepton-quark symmetry, D-V proposes \cite{D16} the exceptional Jordan or \textit{Albert} algebra of $3 \times 3$ hermitian octonionic matrices,
\be
\label{eq17}
J_3^8 = {\cal H}_3 ({\mathbb O}) \, ,
\ee
the only irreducible one which does not admit an associative envelope \cite{A}. Further progress was achieved in \cite{TD,TD-V, DT, T, DT20} by considering the Clifford algebra envelope of its nonexceptional subalgebra $J_2^8$ which fits one generation of fundamental fermions. In these papers, as well as in the present one, we are effectively working with associative algebras which should be viewed as an internal space counterpart of Haag's \textit{field algebra} (see \cite{H}).

\smallskip

A second development, {\it Grand Unified Theory} (GUT), anticipated again in 1973 by Pati and Salam \cite{PS}, became mainstream\footnote{For an enlightening review of the algebra of GUTs and some 40 references see \cite{BH}.} for a time. Fundamental chiral fermions fit the complex spinor representation of $Spin (10)$, introduced as a GUT group by Fritzsch and Minkowski and by Georgi. A preferred symmetry breaking yields the maximal rank semisimple Pati-Salam subgroup,
$$
G_{\rm PS} = \frac{Spin (4) \times Spin (6)}{{\mathbb Z}_2} \subset Spin (10),
$$
\be
\label{eq18}
Spin (4) = SU(2)_L \times SU(2)_R , \ Spin (6) = SU(4) \, .
\ee
We note that $G_{\rm PS}$ is the only GUT group which does not predict a gauge triggered proton decay. It is also encountered in the noncommutative geometry approach to the SM \cite{CCS,BF}. In general, GUTs provide a nice home for the fundamental fermions, as displayed by the two 16-dimensional complex conjugate ``Weyl (semi)spinors'' of $Spin (10)$. Their other representations, however, like the 45-dimensional adjoint representation of $Spin (10)$ are much too big, involve hypothetical particles like leptoquarks which cause difficulties.

\smallskip

 The Clifford algebra\footnote{Aptly called {\it geometric algebra} by its inventor -- see \cite{DL}.} $C\ell_{10}$,  on the other hand, like the Clifford algebra of any even dimensional euclidean vector space, has a unique irreducible representation (IR); in the case of\footnote{For any associative ring ${\mathbb K}$, in particular, for the division rings ${\mathbb K} = {\mathbb R} , {\mathbb C}, {\mathbb H}$, we denote by ${\mathbb K} [m]$  the algebra of $m \times m$ matrices with entries in ${\mathbb K}$.} $C\ell_{10}\cong \mathbb{R}[2^5]$ it is the 32-component real (Majorana) spinor. Viewed as a representation of $Spin(10)$ it splits upon complexification into two 16-dimensional (complex) IRs which can be naturally associated to the left and right chiral fundamental (anti)fermions of one generation:
\be
\label{eq19}
\textbf{32} = \textbf{16}_L + \textbf{16}_R.
\ee
Clifford algebras were also applied to the SM in the 1970's -- see \cite{CG} and references therein. An essential difference in our approach is the use of the octonions with a preferred complex structure in $C\ell_{-6}, C\ell_{8+\nu}$, $\nu = 0,1,2$ restricting the associated gauge group. (Another new point, the use of the ${\mathbb Z}_2$ grading of $C\ell_{10}$ to define the Higgs field, will be discussed in Sect. 1.3 below.)

The Pati-Salam subgroup of $Spin(10)$ is singled out as the stabilizer of the $C\ell_6(\subset C\ell_8\subset C\ell_{10})$ pseudoscalar
$$
\omega_6=\gamma_1...\gamma_6 \,\, \, \mbox {for} \, \, \, \gamma_\alpha=\sigma_1\otimes \epsilon\otimes L_\alpha, \gamma_8=\sigma_1 \otimes\sigma_1 \otimes \un_8 (\in C\ell_{10}), 
$$
\be 
\label{eq110}
 \epsilon=i\sigma_2, \, L_\alpha=L_{e_\alpha}, \, [L_\alpha, L_\beta]_+ = -2\delta_{\alpha \beta}\un_8, \, \alpha, \beta =1, ..., 7.
\ee
Here $L_x$ is the operator of left multiplication in the 8-dimensional real vector space of the octonions, $L_xy=xy$ for $x, y \in \mathbb{O}; \, \sigma_\nu, \nu=0, 1, 2, 3$ are the Pauli matrices ($\sigma_0 = \un_2$). The action of the operators $L_\alpha \in \mathbb{R}[8]$ on the octonion units will be made explicit in Sect. 2.2 (Eq. (\ref{eq29})). The products $\epsilon\otimes L_\alpha, \alpha=1, ..., 7$ and $\sigma_1 \otimes \un_8$ generate the Clifford subalgebra $C\ell_8$ of $C\ell_{10}$. All generators of $C\ell_{10}$ are listed in Eq. (2.23) below. The group $G_{PS}$ (\ref{eq18}) preserves, in fact, each factor in the graded tensor product representation of $C\ell_{10}$:
\be 
\label{eq111}
C\ell_{10} = C\ell_4\hat{\otimes} C\ell_6
\ee
introduced earlier by Furey \cite{F16, F} and exploited in \cite{T21}. The complex structure $J\in SO(10)$ generated by $\omega_6$ will be displayed  and the physical interpretation of the mutually orthogonal projection operators
\be 
\label{eq112} 
\mathcal{P}=\frac{1}{2}(1-i\omega_6), \, \mathcal{P}'= \frac{1}{2}(1+i\omega_6)
\ee
will be revealed in Sect. 2.3.


\subsection{Main message and organization of the paper}  

The present survey focuses on ongoing attempts to answer two questions:

1) Why the arbitrarily looking gauge group of the SM,
\be 
\label{GSM}
G_{SM} = S(U(2)\times U(3)) = \frac{SU(2)\times SU(3)\times U(1)}{\mathbb{Z}_6},
\ee
and what dictates its highly reducible representation for fundamental fermions?

2) How to put together the Higgs field with the gauge bosons? Can we explain their mass ratios?

1. Most physicists accept GUT as an answer to the first question. One has then the intriguing result of Baez and Huerta, \cite{BH}, that $G_{SM}$ appears as the intersection of two popular GUT subgroups of $Spin(10)$: 
$$
G_{SM}=SU(5)\cap G_{PS}\subset Spin(10).
$$
 A top down approach, however, starting with $Spin(10)$ should involve the maximal rank subgroup $U(5)$ instead of $SU(5)$, in line with the philosophy of Borel - de Siebenthal, \cite{BdS}, yielding an extra $U(1)$ factor in the intersection. 
 
The minority trying to go further includes, besides the fans of octonions and the already cited enthusiasts of almost commutative real spectral triples, also Holger Nielsen whose more than two decades of musing over the problem are reviewed in \cite{NB}. Our approach exploits the complex structure and the particle projector $\mathcal{P}$ (\ref{eq112})  associated with the  Clifford pseudoscalar $\omega_6$ (\ref{eq110}). It permeates the entire paper (Sects. 2.3, 3.2-4, 5.1, ...).

2. The second problem has been universally recognized (see, e.g., the popular account \cite{M14}). We follow the superconnection approach anticipated by Ne'eman and Fairlie - for a concise review and references see Sect. 4.1. We exploit the restricted particle projector $\mathcal{P}_r$ which annihilates the sterile neutrino (Sect. 3.3) to deform the Fermi oscillators in the lepton sector into the odd generators of the simple  Lie superalgebra postulated in \cite{N, F79}. The resulting difference between the flavour spaces of leptons and coloured quarks allows to compute the mass ratio $m_H/m_W$ in agreement with experiment (Sect. 4.2). 

The paper aims to be selfcontained and combines in a single narrative our contribution with review of background material. Sect. 2.1 provides a summary of the known triality realization of $Spin(8)$. Sect. 2.2 and Appendix A spell out the relation between left and right multiplication by imaginry octonion units, applied in Sect. 3.4 to display the stabilizer of $R_7$. We should like to single out two messages from Sect. 2.3: (1) the indirect connection between the $C\ell_6$ pseudoscalar and the complex structure $J\in SO(8)$ (\ref{eq221}) (\ref{eq223}); (2) the observation 	that the Lie subalgebra of $so(8)$ that commutes with $\omega_6$ and the electric charge operator $Q$ (\ref{eq233}) is the rank four subalgebra $su(3)_c\oplus u(1)_Q\oplus u(1)_{B-L}$.   Sect. 3.1 contains, along with a glance on the equivalence class of Clifford algebras involving $C\ell(3, 1), C\ell_{-6}(=C\ell(0, 6)), C\ell_{10}$, the observation that the conformal Clifford algebra $C\ell(4, 2)$ of this class is generated by the split octonions and gives rise to their isometry group $SO(4, 4)$. Sect. 3.2 contains one of the main messages of the paper: the SM gauge group (\ref{GSM}) is the soubgroup of $G_{PS}$ (\ref{eq18}) that leaves the sterile neutrino invariant (Proposition 3.1).  Sect. 3.3 discusses superselection rules and the superselection of the weak hypercharge. Sect. 3.4 reviews and comments on recent work \cite{K, FH} on complex structure associated with the right action of $\mathbb{O}$ on the octonion units as well as the derivation of the gauge group for the SM \cite{TD-V} and its left-right symmetric extension \cite{B}.  
 
The Dirac operator $\gamma^\mu (\partial_\mu + A_\mu)$ anticommutes with the chirality $\gamma_5$ and hence intertwines left and right fermions; so does the Higgs field which substitutes a mass term in the fermionic Lagrangian. This has inspired Connes and coworkers, \cite{C, CL, CC}, to identify the Higgs field with the internal space part of the Dirac operator. This idea finds a natural implementation in the Clifford algebra approach to the SM \textit{superconnection}  (reviewed in Sect. 4.1). The concise exposition in Sect. 4 emphasizes our assumptions and some  delicate points, referring the reader for calculational details to the preceding publication \cite{T21}.

We recapitulate our convoluted route to $C\ell_{10}$ in Sect. 5.1. In Sect. 5.2 we compare our solution of the fermion doubling problem with the approach  of \cite{FH1}.  A summary of the main results of the paper is given in Sect. 5.3 which also cites existing (inconclusive) attempts to understand why are there exactly three generations of fundamental fermions.

\section{Triality realization of $Spin (8)$; $C\ell_{-6}$}\label{sec2}
\setcounter{equation}{0}

\subsection{The action of octonions on themselves}

The group $Spin (8)$, the double cover of the orthogonal group $SO(8) = SO({\mathbb O})$, can be defined (see \cite{Br,Y}) as the set of triples $(g_1 , g_2 , g_3) \in SO(8) \times SO(8) \times SO(8)$ such that
\be
\label{eq21}
g_2 (xy) = g_1 (x) \, g_3 (y) \ \mbox{for any} \ x,y \in {\mathbb O} \, .
\ee
If $u$ is a unit octonion, $u^* u = 1$, then the left and right multiplications by $u$ are examples of isometries of ${\mathbb O}$
\be
\label{eq22}
\vert L_u \, x \vert^2 = \langle ux , ux \rangle = \langle x,x \rangle , \ \vert R_u \, x \vert^2 = \langle xu , xu \rangle = \langle x , x \rangle \ \mbox{for} \ \langle u,u \rangle = 1 \, .
\ee
Using the {\it Moufang identity}\footnote{See \cite{S16} for a reader's friendly review of Moufang loops and for a glimpse of the personality of Ruth Moufang (1905-1971).}
\be
\label{eq23}
u(xy)u = (ux)(yu) \ \mbox{for any} \ x,y,u \in {\mathbb O} \, ,
\ee
one verifies that the triple $g_1 = L_u$, $g_2 = L_u \, R_u$, $g_3 = R_u$ satisfies (\ref{eq21}) and hence belongs to $Spin (8)$. It turns out that triples of this type generate $Spin (8)$ (see \cite{Br} or Yokota's book \cite{Y} for a proof).

\smallskip

The mappings $x \to L_x$ and $x \to R_x$ are, of course, not algebra homomorphisms as  {$L_x$} and $R_y$ generate each an associative algebra while the algebra of octonions is non-associative. They do preserve, however, the quadratic relation 
{$xy^* + yx^* = 2 \langle x,y \rangle 1$} (and the octonion conjugation):
\be
\label{eq24}
L_x L_{y^*} + L_y L_{x^*} = 2 \langle x,y \rangle 1 = R_x R_{y^*} + R_y R_{x^*} \,   (L^*_x= L_{x^*}).
\ee
Eq. (\ref{eq24}), applied to the span of the first six imaginary octonion units $e_j$, $j = 1 , \cdots , 6$, setting $L_{e_j} =: L_j$, $R_{e_j} =: R_j$, becomes the defining relation of the Clifford algebra $C\ell_{-6}$:
\be
\label{eq25}
([L_j, L_k]_+:=) L_j L_k + L_k L_j = -2 \delta_{jk} = R_j R_k + R_k R_j , \ j,k = 1 , \cdots , 6 \, .
\ee

\subsection{$C\ell_{-6}$ as a generating algebra of ${\mathbb O}$ and of $so({\mathbb O})$}

The octonions appear in any of the nested Clifford algebras $C\ell_8\subset C\ell_9\subset C\ell_{10}$. In fact, the minimal realization of $\mathbb{O}$ is provided by $C\ell_{-6}$ generated by left multiplication 
$L_\alpha$ by six of the seven imaginary octonion units $e_\alpha$.
In general, $L_x L_y \ne L_{xy}$ (and similarly for $R$), but remarkably, as noted in \cite{F16}, the relation $e_1 (e_2 (e_3 (e_4 (e_5 (e_6 \, e_a)))))) = e_7 \, e_a$ is satisfied for all $a = 1,\cdots , 8 \, (e_8=1)$, so that
\be
\label{eq26}
L_1 L_2 \cdots L_6 = L_{e_7} =: L_7 , \ R_1 R_2 \cdots R_6 = R_{e_7} =: R_7 \, .
\ee
While {$L_x R_x = R_x L_x$ (for $x \in {\mathbb O}$}) the non-associativity of the algebra of octonions is reflected in the fact that for $x \ne y$, $L_x$ and $R_y$, in general, do not commute. The Lie algebra $so(8)$ is spanned by the elements of negative square of $C\ell_{-6}$. If we denote the exterior algebra on the span of $L_1 , \cdots , L_6$ by
$$
\Lambda^* \equiv \Lambda^* C\ell_{-6} = \Lambda^0 + \Lambda^1 + \cdots + \Lambda^6 \ \Bigl(\Lambda^1 = \underset{1 \leq j \leq 6}{\rm Span} \, L_j , \ \Lambda^6 = \{ {\mathbb R} \, L_7 \}\Bigl)
$$
then $so(8) = \Lambda^1 + \Lambda^2 + \Lambda^5 + \Lambda^6$. (Accordingly, the 28-dimensional adjoint representation of $so(8)$ splits into four irreducible representations of $so(6): \textbf{28}=\textbf{6}+\textbf{15}+\textbf{6}^*+\textbf{1}$. In particular, $\Lambda^5=Span\{L_{\alpha 7}, \alpha=1, ..., 6\}$, for $L_{\alpha \beta}$ defined below.) A basis of the Lie algebra, given by
\be
\label{eq27}
L_{\alpha 8} = \textstyle\frac12 \, L_{\alpha} , \ L_{\alpha\beta} = - \textstyle\frac14 \, [L_{\alpha} , L_{\beta}] , \ \alpha , \beta = 1 , \cdots , 7
\ee
obeys the standard commutation relations (CRs) for $so(n)$ (here $n=8$):
$$
[L_{ab} , L_{cd}] = \delta_{bc} \, L_{ad} - \delta_{bd} \, L_{ac} + \delta_{ad} \, L_{bc} - \delta_{ac} \, L_{bd} , 
$$
\be
\label{eq28}
L_{ab} = \textstyle\frac14 (L_a L_b^* - L_b L_a^*), \ a,b,c,d = 1,2,\cdots , 8 \,  (L^*_\alpha=-L_\alpha, L^*_8=L_8)
\ee
(and similarly for $R_{ab}$). Each element of $so(8)$ of square $-1$ defines a {\it complex structure} (see Sect. 2.3). Following \cite{FH} we shall single out the {\it Clifford pseudoscalars} $L_7$ and $R_7$ (\ref{eq26}) (called {\it volume forms} in the highly informative lectures \cite{M} and Coxeter elements in \cite{T11}). We shall use the (mod 7) multiplication rules of \cite{B02} for the imaginary octonion units
$$
L_i e_j (= e_i e_j) = - \delta_{ij} + f_{ijk} \, e_k , \ f_{ijk} = 1 \ 
$$
\be
\label{eq29}
\mbox{for} \ (i,j,k) = (1,2,4) (2,3,5) (3,4,6) (4,5,7) (5,6,1) (6,7,2) (7,1,3)
\ee
and $f_{ijk}$ is fully antisymmetric within each of the above seven triples. The Clifford pseudoscalar is naturally associated with the Cartan subalgebra of $so(6)$ spanned by
\be
\label{eq210}
(L_{13} , L_{26} , L_{45}) \ \mbox{as} \ L_7 (e_1 , e_2 , e_4) = (e_3 , e_6 , e_5) \, .
\ee
We can write
\be
\label{eq211}
L_7 = 2^3 L_{13} L_{26} L_{45} \ (\mbox{as} \ 2L_{13} = L_1 L_3^* = -L_1 L_3 \ \mbox{etc.}).
\ee

The infinitesimal counterpart of (\ref{eq21}) reads
$$
T_{\alpha} (xy) = (L_{\alpha} \, x)y + x(R_{\alpha} \, y) \ \mbox{for} \ \alpha , x , y \in {\mathbb O} , \ \alpha^* = - \alpha \, ,
$$
\be
\label{eq212}
\mbox{i.e.} \qquad T_{\alpha} = L_{\alpha} + R_{\alpha} \, .
\ee

\noindent There is an involutive outer automorphism $\pi$ of the Lie algebra $so(8)$ such that
\be
\label{eq213}
\pi (L_{\alpha}) = T_{\alpha} , \ \pi (R_{\alpha}) = - R_{\alpha} , \ \pi (T_{\alpha}) = L_{\alpha} \ (\pi^2 = id) \, .
\ee
As proven in Appendix A
\be
\label{eq214}
\pi (L_{ab}) = E_{ab} \ \mbox{where} \ E_{ab} \, e_c = \delta_{bc} \, e_a - \delta_{ac} \, e_b \ (a,b,c = 1,2,\cdots ,8, \ e_8 = 1).
\ee
$(L_{ab}), (E_{ab})$ and $(R_{ab})$ provide three bases of $so(8)$, each obeying the CRs (\ref{eq28}). They are expressed by each other in terms of the involution $\pi$:
\be
\label{eq215}
L_{ab} = \pi (E_{ab}) , \ E_{\alpha 8} = L_{\alpha 8} + R_{\alpha 8} , \ \alpha = 1 , \cdots , 7 \, .
\ee
We find, in particular -- see Appendix A:
$$
 L_7 = 2L_{78} = E_{78} - E_{13} - E_{26} - E_{45}=2E_{78}-R_7; \, 2L_{13}=E_{13}-E_{26}-E_{45}-E_{78}, 
$$
\be
\label{eq216}
 2L_{26}= E_{26}-E_{13} - E_{45} E_{78},  \, 2L_{45} = E_{45} - E_{13}- E_{26} - E_{78}.
\ee
While $L_{78} = 4L_{13} L_{26} L_{45}$ (\ref{eq211}) commutes with the entire Lie algebra $spin (6) = su(4)$ the $u(1)$ generator  (whose physical meaning is revealed by (2.32)) 
\be
\label{eq217}
C_1 = L_{13} + L_{26} + L_{45} \ \mbox{centralizes} \ u(3) = u(1) \oplus su(3) \subset su(4)
\ee
where the second summand is the unbroken colour Lie algebra $su(3)=su(3)_c$.

\subsection{Complex structure and symmetry breaking in $C\ell_n$}

The algebra $C\ell_8$ is generated by two-by-two hermitian matrices whose elements involve the operators $L_a$ of left multiplication by octonion units:
\be 
\label{gammaCl8}
\gamma_a=\begin{pmatrix} 0 &L_a \\ L_a^* &0 \end{pmatrix}, \, L_a=L_{e_a}, \, L_a^*=L_{e_a^*}, \, a=1, ..., 8.
\ee
Here $e_8=1(=e_8^*), L_8=\un_8$ is the unit operator in $\mathbb{R}^8$; $L_\alpha^*=-L_\alpha$ for $\alpha=1,..., 7$ so that the $C\ell_{10}$ generators $\gamma_\alpha$ (\ref{eq110}) are obtained from the above (for $\alpha=1, ..., 7)$ by tensoring with the $2\times 2$ unit matrix $\sigma_0$. 

A compact way to identify the particle states in a Clifford algebra $C\ell_{2n}$ is to introduce a complex structure which as we shall demonstrate gives rise to a fermionic Fock space in $C\ell_{2n}$.

\smallskip

{\footnotesize A \textit{complex structure}  in an even-dimensional real euclidean space $E_{2n}$ with a positive definite symmetric scalar product $(X, Y) = (Y,X)$ is an orthogonal transformation $J$ of $E_{2n}$ of square -1; equivalently, a complex structure is a skewsymmetric isometry $J$ of $E_{2n}$:
\be 
\label{Jskew}
(JX, JY)=(X, Y), \, (JX, Y)=-(X, JY),  \, \, \forall X, Y \in E_{2n}.
\ee
For a non zero vector $X$ and a complex structure $J$ the vector $Y=JX$ is orthogonal to $X$ (and has the same norm):
$$
Y=JX \Rightarrow (X, Y) = 0  \, \, ((X, X) = (Y, Y)>0). 
$$
It follows that for each complex structure $J$ in $E_{2n}$ there exists an orthonormal basis  of the form $(\gamma_1, ..., \gamma_n, J\gamma_1, ..., J\gamma_n)$ in $C\ell_{2n}$. Then $a_j=\frac{1}{2}(\gamma_j-iJ\gamma_j)$ and $a_j^*=\frac{1}{2}(\gamma_j+iJ\gamma_j), j=1, ..., n$ span each the image in $C\ell_{2n}(=C\ell_{2n}(\mathbb{C}))$  of a maximal isotropic subspace of the complexification of $E_{2n}$. Together they yield a realization of the \textit{canonical anticommutation relations} (CAR).  Fermionic oscillators have been used in the present context in \cite{B77, F}.    The complex structure in $so(2n)$ involves a distinguished maximal (rank $n$) Lie subalgebra (a notion studied in \cite{BdS}), $u(n)\subset so(2n)$, generated by the products $a_ja_k^*$.  It also selects two distinguished $u(n)$ singlet states in $C\ell_{2n}$, the \textit{vacuum}, annihilated by all $a_j$ and its antipode, annihilated by the $a_j^*$. Both singlets are annihilated by the simple part $su(n)$ of $u(n)$. 

Complex structures have been studied in relation to spinors by \'Elie Cartan (since 1908), Veblen (1933), Chevalley (1954). For
a carefully written servey with historical highlights - see \cite{BT}. The second reference [BT] and the concise modern exposition \cite{D}  connect simple spinors to the states in a fermionic Fock space. We have been also influenced by their use (in  $SO(9)$) by Krasnov \cite{K} and by  relating them to Clifford pseudoscalars in \cite{FH}.}

The pseudoscalar $\omega_6$ of $C\ell_6$ belongs to $C\ell_8$ but only defines a complex structure through its action on the octonion units. More precisely, taking the basic relations (\ref{eq29}) and the identity 
$\epsilon^2=-\sigma_0$ into account, we can write 
$$
\omega_6=-\sigma_0\otimes L_7 \, \mbox{where}\, L_7e_a=-\sum_b J_{ab}e_b, J_{\alpha\beta} = f_{7\alpha\beta}, \alpha, \beta=1 ,..., 6,
$$
\be
\label{eq221}
J_{ab}=-J_{ba}, \, J_{13} = J_{26} = J_{45} = -J_{78}= -1.
\ee
{\footnotesize Warning: due to nonassociativity of $\mathbb{O}, \, -L_7e_3=e_1$ does not imply $-L_7L_3=L_1$ etc.}

We shall see that each of the subgroups $Spin(n)^{\omega_6}$ of the spin groups of $C\ell_n$ for $n=8, 9, 10$ that leaves $\omega_6$ invariant is relevant for particle physics:
\be   
\label{8-9-10}
Spin(8)^{\omega_6} = U(4), \, \, Spin(9)^{\omega_6}=SU(4)\times SU(2), \,\, Spin(10)^{\omega_6}=G_{PS}.
\ee
The $U(1)$ factor in the $U(4)$ of $Spin(8)^{\omega_6}$ and the $SU(2)$ in $Spin(9)^{\omega_6}$ are generated by the third, respectively by all three components of the "total weak isospin" $\textbf{I}=\textbf{I}^L + \textbf{I}^R$  as will be made explicit in Sect. 3.2.

We shall define \textit{a complex structure $J$ corresponding to $\omega_6$} by extrapolating (\ref{eq221}) to a transformation of the gamma matrices of $C\ell_8$:
\be 
\label{eq223}
J: \gamma_a \rightarrow \sum_bJ_{ab}\gamma_b=\omega_J\gamma_a\omega_J^*, \, \omega_J=\frac{1}{4}(1+\gamma_{13})(1+\gamma_{26})(1+\gamma_{45})(1-\gamma_{78}).
\ee

We can extend the basis (\ref{gammaCl8}) of $C\ell_8$ to $C\ell_{10}$ setting (cf. (\ref{eq110}))
$$
\gamma_\alpha = \sigma_1\otimes \epsilon\otimes L_\alpha \, \mbox{for} \, \alpha = 1, ..., 7,
$$
\be 
\label{eq224}
\gamma_8=\sigma_1\otimes \sigma_1\otimes \un_8, \, \gamma_9=\sigma_2\otimes \sigma_0\otimes \un_8, \, \gamma_{10}=\sigma_1\otimes \sigma_3\otimes \un_8.
\ee
 In particular, $C\ell_9 = \mathbb{R}[16] \oplus \mathbb{R}[16]$ is generated by the $32\times 32$ matrices $\gamma_1, ..., \gamma_9$ which commute with $\omega_9=\gamma_1\gamma_2...\gamma_9=\sigma_2\otimes \sigma_3\otimes \un_{8}$. The Lie subalgebra of $so(n)$ of dervations of $C\ell_n, n=8, 9, 10$ which commute with $\omega_6$ (\ref{eq110}) is $so(6) \oplus so(n-6)$. For $n=10$ it is the Lie algebra $\mathfrak{g}_{PS}$ of the Pati-Salam group (\ref{eq18}) that respects the tensor product representation (\ref{eq111}) of $C\ell_{10}$. 

We now proceed to give meaning to the projection operator
\be
\label{eq225}
{\cal P} = \frac{1-i\omega_6}2 \quad ({\cal P}^2 = {\cal P}) , \ tr \, {\cal P} = tr(1-{\cal P})= 2^{\ell-1} \, \mbox{for} \,  n=2\ell(=6,  8, 10).
\ee
To begin with we introduce the isotropic elements  \footnote{It would be natural to use a    \textit{chiral basis} in which the matrices $i\gamma_\beta, \beta=3, 6, 5, 7, 9$ are real  skewsymmetric, such that the $so(10)$ Cartan generators $i\textsl{\textsl{•}}\gamma_{\alpha \beta}, (\alpha, \beta)=(1,3), (2, 6), (4, 5), (7, 8), (9, 10)$ are real diagonal.} 
\be 
\label{eq226}
2b_1=(1-iJ)\gamma_1=\gamma_1+i\gamma_3, \, 2b_2=(1-iJ)\gamma_2=\gamma_2+i\gamma_6, 2b_3=(1-iJ)\gamma_4=\gamma_4+i\gamma_5,
\ee
which correspond to  the projected octonion units $\frac{1}{2}(1+iL_7)e_\ell, \, \ell=1, 2, 4$. Together with their conjugates $b_j^* (b_1^* = \frac{1}{2}(\gamma_1-i\gamma_3)$ etc.) they realize the CAR
\be 
\label{eq227}
[b_j, b_k]_+=0, \, [b_j, b_k^*]_+=\delta_{jk}, \, j, k=1, 2, 3.
\ee
The annihilation operators $b_j$ span the (maximal) 3-dimensional isotropic subspace ${\cal H}^{(1,0)}$ of the 6-dimensional complex vector space $\mathbb{C}E_6$ while $b_j^*$ span its orthogonal complement ${\cal H}^{(0,1)}$; we have:
\be 
\label{eq228} 
J (1\mp iJ)\gamma_\ell = \pm i (1\mp iJ)\gamma_\ell \, \, \mbox{for} \, \, \ell=1, 2, 4.
\ee
The commuting hermitian elements $i\gamma_{\ell 3\ell(mod 7)}, \, \ell=1, 2, 4$, which span a Cartan subalgebra of the comlexified $so(6)$, can be expressed as commutators of $b_j^*$ and $b_j, j=1, 2, 3$ or as differences of the associated projection operators $p'_j-p_j$:
$$
i\gamma_{\ell 3\ell}=[b_\ell^*, b_\ell]=p'_\ell - p_\ell,  \, \ell=1, 2, \, \, i\gamma_{45}=[b_3^*, b_3]= p'_3 - p_3,
$$
\be
\label{eq229}
p_j=b_jb_j^*, p'_j=b_j^*b_j=1-p_j, p_jp'_j=0, \,\, j=1, 2, 3.
\ee
In terms of these operators the $C\ell_6$ pseudoscalar and the projector $\cal{P}$ assume the form:
$$ 
i\omega_6=(p'_1-p_1)(p'_2-p_2)(p'_3-p_3)={\cal P}'-{\cal P}, \, \mathcal{P}' = 1 - \mathcal{P},
$$
\be 
\label{eq230}
{\cal P}=\ell + q, \, \ell=p_1 p_2 p_3, \, q=q_1+q_2+q_3, \, q_j = p_j p'_k p'_\ell, 
\ee
the triple $(j, k, \ell)$ being a permutation of $(1, 2, 3)$.

We shall identify the generators (of the comlexification $s\ell(3, \mathbb{C})$) of $su(3)$ with the traceless part of the matrix $(b_j b_k^*)$ whose elements belong to $\mathcal{H}^{(1,1)}$. Then the splitting (2.29) of $\cal{P}$ into the $su(3)$ singlet $\ell$ and the triplet $q$ implements the lepton-quark splitting anticipated by its image (1.1) on the octonions. We shall thus interpret the 1-dimensional projectors $\ell$ and $q_j$ as describing the lepton and the coloured quark \textit{states} in $C\ell_6$. The states $\ell$ and $q_j$ are mutually orthogonal idempotents, $\ell$ playing the role of Fock vacuum in $C\ell_6$:
\be 
\label{eq231}
\ell^2=\ell, \, \ell q_j=0, \, q_jq_k=\delta_{jk}q_j, \, \, b_j\ell=0=\ell b_k^*.
\ee

\textbf{Remark.} - We shall argue in Sect. 3.3 that the identification of $\mathcal{P}$ as a particle subspace projector (adopted in \cite{DT20}) would be only justified if we have a clear distinction between particles and antiparticles. This can be claimed for the 30 fundamental (anti)fermions of the $C\ell_{10}$ multiplet $\textbf{32}$ (\ref{eq19}) which have different quantum numbers with respect to the gauge Lie algebra $\mathfrak{g}_{SM}$ of the SM. It fails in the 2-dimensional subspace of sterile neutrinos annihilated by $\mathfrak{g}_{SM}$; $\nu_R$ and $\bar{\nu}_L$ are allowed to form a coherent superposition - a Majorana spinor. We adopt in Sects. 3.3 and 4. the restricted projector $\ell_r$ (3.21) on the (3- rather than 4-dimensional) lepton subspace, excluding the sterile neutrino.

In order to extend the Fock space picture to $C\ell_8$ we shall set
\be 
\label{eq232}
 i\gamma_7=a^*-a, \, \gamma_8=a+a^* \, \Rightarrow i\gamma_{78}=[a^*, a]
 \ee
 where the pair $(a^*, a)$ describes another Fermi  oscillator $([a, a^*]_+=1)$ anticommuting with $b_j, b_k^*$. We shall fix the physical interpretation of $[a^*, a]$ by postulating that the \textit{electric charge operator} is given by
$$
Q:= \frac{1}{3}\sum_{j=1}^3 b_j^*b_j-a^*a = \frac{1}{2}( B-L -  [a^*, a]),
$$  
\be 
\label{eq233}
\mbox{where} \, \, B-L = \frac{2i}3 \, (L_{13} + L_{26} + L_{45})  = \frac{1}3\sum_j [b^*_j, b_j] ,
\ee
stands for the difference between the baryon and the lepton numbers. $B-L$ takes eigenvalues $\pm \frac13$ for (anti)quarks and $\mp 1$ for (anti)leptons. Demanding that the gauge Lie algebra within $so(8)$ commutes with both $\omega_6$ and $Q$ we shall further reduce it from $so(6)\oplus so(2)$ to the rank four Lie subalgebra  
\be
\label{eq234}
\mathfrak{g}_4 = su(3)_c\oplus u(1)_Q \oplus u(1)_{B-L} = \{X\in u(4), [X,Q]=0\} . \,
\ee
The knowledge of the  charges $Q, B-L$ along with the colour Lie algebra allows to identify the primitive idempotents of $C\ell_8$, given by $\ell, q_j$ multiplied by $aa^*$ or $a^*a$, with the fundamental femions:
\be 
\label{eq235}
\ell aa^*=\nu, \, \ell a^*a=e, \, \, \, q_jaa^*=u_j, \, q_ja^*a=d_j. \,
\ee
The "isotopic doublets" $(\nu, e)$ and $(u_j, d_j)$ stand for neutrino / electron and up / down coloured quarks. We see, in particular, that the Fock vacuum in $C\ell_8$ associated with the complex structure (\ref{eq221}) is identified with the         neutrino (as it has no charge and $a\nu=0=b_j\nu$). Note that the subalgebra of $\mathfrak{g}_4$ which annihilates $\nu$ is the known unbroken gauge Lie algebra $u(3)$ of the SM:
\be 
\label{eq236}
u(3)_{SM} =  su(3)_c \oplus u(1)_Q = \{X\in\mathfrak{g}_4; X\nu = 0\} . \,
\ee
This picture ignores chirality which will find its place in $C\ell_{10}$ (Sect. 3.2).  

\bigskip

\section{ The internal space subalgebra of $C\ell_{10}$}\label{sec3}
\setcounter{equation}{0}

\subsection{Equivalence class of Lorentz like Clifford algebras}

Nature appears to select real Clifford algebras $C\ell (s,t)$ of the equivalence class of $C\ell (3,1)$ (with Lorentz signature in four dimensions) in \'Elie Cartan's classification (which involves \footnote{ The 10-fold classification of $\mathbb{Z}_2$ graded Clifford algebras also involves signs coming from squaring two antiunitary charge conjugation operators - see \cite{M} Chapter 13, pp. 87-125.} the signs, $\omega^2(s, t)$ and $(-1)^{s-t}$):
\be
\label{eq31}
C\ell (s,t) = {\mathbb R} [2^n] , \ \mbox{for} \ s-t = 2 ({\rm mod} \, 8) , \ s+t = 2n \, .
\ee
The elements of $C\ell(s, t)$ are operators acting in the $2^n$ dimensional real vector space $\mathcal{S}$ of \textit{Majorana spinors}. The space $\mathcal{S}$ admits no nontrivial {\it real} $Spin(s, t)$ invariant subspace. If $\gamma_1 , \cdots , \gamma_{2n}$ is the image in $C\ell(s,t)$ of an orthonormal basis of the underlying vector space ${\mathbb R}^{s,t}$ then the Clifford volume form 
\be
\label{eq32}
\omega(s,t) = \gamma_1 \cdots \gamma_{2n} , \ 2n = s+t , \ \omega(s,t)^2 =(-1)^{\binom{s-t}{2}} = -1 \,\mbox{for} \, s-t=2 (mod 8),
\ee
defines a complex structure which commutes with the action of $Spin (s,t)$. For a (complex) \textit{chiral basis} in which $\omega(s,t)$ is diagonal (and hence pure imaginary for $s-t=2 (mod 8)$) the {\it Dirac spinor} splits into two $2^{n-1}$-dimensional complex\textit{ Weyl (semi) spinors} transforming under {\it inequivalent complex conjugate} representations of  $Spin(s,t)$. The corresponding projectors $\Pi_L$ and $\Pi_R$ on left and right spinors are given in terms of the chirality $\chi$ which coinsides with $\gamma_5$ for $C\ell(3, 1)$:
$$
\Pi_L = \textstyle\frac12 (1-\chi) , \ \Pi_R = \textstyle\frac12 (1+\chi) , \ \chi = i \omega(s,t) \, ,
$$
\be
\label{eq33}
\chi^2 = \un \Leftrightarrow \Pi_L^2 = \Pi_L , \ \Pi_R^2 = \Pi_R , \ \Pi_L \Pi_R = 0 , \ \Pi_L + \Pi_R = \un \, .
\ee

Another interesting example of the same equivalence class (also with indefinite metric) is the {\it conformal Clifford algebra} $C\ell (4,2)$ (with isometry group $O(4,2)$). We shall demonstrate that just as $C\ell_{-6}$ was viewed (in Sect.~2.2) as the {\it Clifford algebra of the octonions}, $C\ell (4,2)$ plays the role of the {\it Clifford algebra of the split octonions}  (also appearing in bi-twistor theory, \cite{P23}):
$$
 x = v + V + l (w+W), \ v,w \in {\mathbb R} , \ V = iV_1 + jV_2 + kV_3 , \ W = iW_1 + jW_2 + kW_3 
$$
\be
\label{eq34}
i^2 = j^2 = k^2 = ijk = -1 ,  l^2 = 1 , \ V l =-l V \, .
\ee 
Indeed, defining the mapping (cf. (\ref{eq16}))
$$
i \to \gamma_{-1} , \ j \to \gamma_0 , \ l \to \gamma_1 , \ jl \to \gamma_2 , \ \ell k \to \gamma_3 , \ \ell i \to \gamma_4
$$
\be
\label{eq35}
[\gamma_{\mu} , \gamma_{\nu}]_+ = 2\eta_{\mu\nu} \un , \ \eta_{11} = \eta_{22} = \eta_{33} = \eta_{44} = 1 = -\eta_{-1,-1} = -\eta_{00} \
\ee 
we find that the missing split-octonion (originally, quaternion) imaginary unit $k$ ($= ij = -ji$) can be identified with the $C\ell (4,2)$ pseudoscalar:
\be
\label{eq36}
\omega(4,2) = \gamma_{-1} \, \gamma_0 \, \gamma_1 \, \gamma_2 \, \gamma_3 \, \gamma_4 \leftrightarrow k , \ \omega(4,2)^2 = -1 , \ [w(4,2) , \gamma_{\nu}]_+ = 0 \, .
\ee
The conjugate to the split octonion $x$ (\ref{eq34}) and its norm are
$$
 x^* = v - V - l (w+W) , \ N(x) = xx^* = v^2 + V^2 - w^2 - W^2, 
$$
so that the isometry group of $\widetilde{\mathbb O}$ is $O(4,4)$. (In particular, the maximal compact subalgebra $so(4) \oplus so(4)\subset so(4, 4)$ is spanned by $\gamma_{jk}, j, k=1,...,4$ and by $\omega_{4,2}, \gamma_\alpha, \alpha=-1, 0$, and their commutators. The remaining 16 noncompact generators of $so(4, 4)$ involve the square-one matrices $\gamma_j, \gamma_\alpha\gamma_j, \gamma_j\omega_{4,2}$.)

\smallskip

As we are interested in the geometry of the internal space of the SM, acted upon by a compact gauge group we shall work with (positive or negative) definite Clifford algebras $C\ell_{2\ell}$, $\ell = 1 ({\rm mod} \, 4)$. The algebra $C\ell_{-6}$, considered in Sect.~2, belongs to this family (with $\ell = -3$). For $\ell =1$ we obtain the Clifford algebra of 2-dimensional conformal field theory; the 1-dimensional Weyl spinors correspond to analytic and antianalytic functions. Here we shall argue that for the next allowed value, $\ell = 5$, the algebra $C\ell_{10} = C\ell_4 \, \widehat\otimes \, C\ell_6$ (\ref{eq111}), fits beautifully the internal space of the SM, if we associate the two factors to colour and flavour degrees of freedom, respectively. We shall strongly restrict the physical interpretation of the generators $\gamma_{ab} \left(=\frac12 \, [\gamma_a , \gamma_b ] , \ a,b = 1 , \cdots , 10 \right)$ of the derivations of $C\ell_{10}$ by demanding that the splitting (\ref{eq111}) of $C\ell_{10}$ into $C\ell_4$ and $C\ell_6$ is preserved. This amounts to select a first step of symmetry breakings of the GUT group $Spin (10)$ leading to the semisimple Pati-Salam group $(Spin (4) \times Spin (6))/{\mathbb Z}_2$ (\ref{eq18}). Each summand, $so(4)$ and $so(6)$, of $\mathfrak{g}_{PS}$, expressed in terms of Fermi creation and annihilation operators, has a distinguished Lie subalgebra, $u(2)$ respectively $u(3)$, which belongs to $\mathcal{H}^{1,1}$. We identify the leptons and quarks with $u(3)$ singlets and triplets. This identification implements the lepton-quark symmetry alluded to by (\ref{eq11}). 

\subsection{$\mathfrak{g}_{SM}$ as annihilator of sterile  neutrino}

We proceed to extend the complex structure $J$ (\ref{eq221}), (\ref{eq223}) to $C\ell_{10}$, expressing, in particular, the electroweak gauge group generators in terms of the fermionic oscillators corresponding to the $C\ell_4$ factor in (\ref{eq111}).  
To this end we complement the definition (\ref{eq226}) of $b_j$ by 
\be 
\label{eq37}
2a_1=(1-iJ)\gamma_{10}=\gamma_{10}-i\gamma_9, \, 2a_2=\gamma_8-i\gamma_7 \Rightarrow i\gamma_{78}=[a_2^*, a_2], i\gamma_{9 10}=[a_1^*, a_1]
\ee
(where $\gamma_a$ are given by (\ref{eq224})). In particular, $(a_2, a_2^*)$ coincide with the unique flavour fermionic  oscillator $(a,  a^*)$ (\ref{eq232}) of $C\ell_8$. They allow to define two pairs of complementary projectors
\be
\label{eq38}
\pi_{\alpha} = a_{\alpha} a_{\alpha}^* , \ \pi'_{\alpha} = a_{\alpha}^* a_{\alpha} = 1 - \pi_{\alpha} , \ \alpha = 1,2 , \,
\pi_\alpha\pi'_\alpha=0, \, \pi_\alpha + \pi'_\alpha=1.
\ee
The three pairs of colour $(p_j, p'_j, j=1, 2,3$) and two pairs of flavour ($\pi_\alpha, \pi'_\alpha, \alpha=1,2$) projectors give rise to a ($2^5=32$-dimensional) maximal abelian subalgebra of $C\ell_{10}$ of commuting observables. The flavour gauge Lie algebra gnerators,  the left and right chiral isosspin components, are expressed in terms of $a_\alpha^{(*)}$:
$$
I_+^L = a_1^* \, a_2 , \ I_-^L = a_2^* \, a_1 , \ [I_+^L , I_-^L] = 2I_3^L = \pi'_1 \pi_2 - \pi_1 \pi'_2 = \pi'_1 - \pi'_2 \, ;
$$
\be
\label{eq39}
I_+^R = a_1 \, a_2 , \ I_-^R = a_2^* \, a_1^* , \ [I_+^R , I_-^R] = 2 I_3^R = \pi_1 \pi_2 - \pi'_1 \pi'_2 = \pi_2 - \pi'_1 \, .
\ee
The \textit{chirality operator} $\chi=\Pi_R-\Pi_L$ is expressed in terms of the $C\ell_{10}$ pseudoscalar (as implied by (\ref{eq33}) for $s=10, t=0$):
$$
\chi =  i\omega_{10}=i\omega_6\gamma_{78}\gamma_{9 10}=(\mathcal{P}'-\mathcal{P})[a_1^*, a_1][a_2, a_2^*]\, = (\mathcal{P}' - \mathcal{P})(P_1 - P'_1),
$$
\be 
\label{eq310}
 \, P_1=(2I^L_3)^2=\pi'_1\pi_2+\pi_1\pi'_2 , \, P'_1 = (2I_3^R)^2 = \pi_1 \pi_2 + \pi'_1\pi'_2 = 1-P_1,
\ee
so that $\Pi_L=\mathcal{P}P_1 + \mathcal{P}'P'_1, \, \Pi_R = \mathcal{P}P'_1+\mathcal{P}'P_1$. Within the particle subspace $\mathcal{P}$ the operator $P_1$ projects on left chiral, $P'_1$ on right chiral fermions.

The sum $I_3 = I_3^L + I_3^R$ coincides with the total isospin projection that generates the commutant $u(1)$ of $su(4)$ in $u(4)$ - see the discussion after Eq. (\ref{8-9-10}) in Sect. 2.3.  Conversely, $I_3^L, I_3^R$ appear as chiral projections of $I_3$:
$$
2I_3=2I_3^L+2I_3^R=[a_2, a_2^*] = \pi_2-\pi'_2 = 2Q - (B-L),
$$
\be 
\label{eq311}   
(2I_3)^2 = 1, \,  I_3^L=P_1 I_3 P_1, \, I_3^R =  P'_1I_3 P'_1 \, \, (I_3^L I_3^R = 0).
 \ee
The identification of the vacuum vector, $a_1a_2b_1b_2b_3$ (annihilated by all $a_\alpha, b_j$) becomes consequential if we demand that this ket-vector is a singlet with respect to the gauge group of the SM. The fact that the left and right isospin cannot vanish simultaneously (since $(2(I_3^L+I_3^R))^2=1$) implies that the Lie algebra $\mathfrak{g}_{SM}$ of the SM should be chiral:
\be 
\label{312} 
\mathfrak{g}_{SM}\subset u(2)\oplus u(3), \mbox{where} \,  u(2)= su(2)_L\oplus u(1)_{I_3^R}, \, u(3)=su(3)_c\oplus u(1)_{B-L}.
\ee
It is therefore rewarding that we can identify the Fock space vacuum in $C\ell_{10}$ (given by $\nu$ of (\ref{eq235}) for $C\ell_8$) with the (right handed, hypothetical) \textit{sterile neutrino} (in fact, $\nu_R$ and its antipode $\bar{\nu}_L$ do not interact with the gauge bosons):
\be 
\label{313}
 \nu_R = \pi_1\pi_2\ell, \, a_\alpha \nu_R =0 (= \nu_R a_\alpha^*), \, b_j\nu_R=0 \Leftrightarrow \bar{\nu}_L= \pi'_1\pi'_2\ell', \, a_\alpha^*\bar{\nu}_L=0 \, \mbox{etc}.
\ee

The role of the electric charge $Q$ (2.33) which breaks the $u(4)$ symmetry of $\omega_6$ in $so(8)$ to $u(3)\oplus u(1)_Q$ is played by the \textit{weak hypercharge} $Y$ in $so(10)$:
\be 
\label{eq314}
\frac12 Y = \frac13 \sum_{j=1}^3 b_j^* b_j - \frac12 \sum_{\alpha = 1}^2 a_{\alpha}^* a_{\alpha} = \frac12 \sum_{\alpha = 1}^2 a_{\alpha} a_{\alpha}^* - \frac13 \sum_{j=1}^3 b_j b_j^*.
\ee
They both annihilate the respective vacuum state as well as its antipode. This is made obvious by the two forms of $Y$ in  Eq. (\ref{eq314}) as sums of normal and of antinormal products.  By definition $Y$ belongs to the centre of the broken symmetry subalgebra of $\mathfrak{g}_{PS}$. As pointed out in \cite{T21} - and will be discussed below in Sect. 3.3 - it gives rise to a superselection rule in the SM.  
 
 The significance of choosing the sterile neutrino as a Fock vacuum is summarized by the following
 
 \textit{Proposition 3.1} The Lie subgroup of $G_{PS}$ (\ref{eq18}) that leaves the Fock vacuum $\nu_R$ (3.13) invariant is the SM gauge group (1.13).  

\textit{Proof}. We shall first complete the argument that the maximal Lie subalgebra of $\mathfrak{g}_{PS}$ annihilating the sterile neutrino is $\mathfrak{g}_{SM}$.  We have already noted that the Lie subalgebra of $\mathfrak{g}_{PS}$ for which the vacuum transforms as a singlet is $u(2)\oplus u(3)$ (\ref{312}). This follows from the observation that generators involving $a_1^*a_2^*$ and $b_j^*b_k^*$ transform $\nu_R$ into a righthanded electron $e_R$ and an up quark $u_R$, respectively. It remains to analyze the 2-dimensional centre $u(1)_{B-L}+u(1)_{I_3^R}$ of this extended algebra. $\nu_R$ and $\bar{\nu}_L$ are eigenvectors of both generators with eigenvalues of opposite sign; only multiples of $Y$ annihilate the sterile neutrino:
\be 
\label{eq315}
(2I_3^R - 1)\nu_R =0 =(B-L+1)\nu_R, \, Y=B-L+2I_3^R \Rightarrow Y\nu_R = 0 = Y\bar{\nu}_L.
\ee 
This establishes the characterization of the Lie algebra $\mathfrak{g}_{SM}$ as annihilator of sterile neutrino. It will be straightforward to extend the result to the SM gauge group (\ref{GSM}) after displaying the quantum numbers of the fundamental fermions in the following subsection. 
 

\bigskip

\subsection{Superselection rules. Restricted particle subspace}

The weak hypercharge (\ref{eq314}) (\ref{eq315}) generates the $u(1)$ centre of the gauge Lie algebra of the SM, hence commutes with all gauge transformations. It is conserved not only in the observed micro processes but even in hypothetical ones, like a possible proton decay (with a conserved $B-L$), or in the presence of a Majorana neutrino (a coherent superposition of $\nu_R$ and $\bar{\nu}_L$) that would break $B-L$ by two units. It was proposed in \cite{T21} as a \textit{superselection rule}, assuming that $Y$ commutes with all observables. The Jordan algebra of the 32-dimensional space of internal observables of one generation splits into 11 superselection sectors corresponding to the 11 different eigenvalues of $Y$ (see Appendix to \cite{T21}).

{\footnotesize Superselection rules (SSR) were introduced by Wick, Wightman, Wigner \cite{WWW} in 1952. The superselection of the electric charge has been thoroughly discussed (see the second paper in \cite{WWW} and the review \cite{G07}); for more references and a historical survey addressed to philosophers - see\cite{ E}. The charge  Q (\ref{eq233}) is superselected by the exact symmetry of the SM (otherwise $I_{\pm}^L$ do not commute with it). SSRs are also related to measurement theory, \cite{T11}. SSR and superselection sectors are an essential part of the  Doplicher-Haag-Roberts
reconstruction of quantum fields from the algebra of observables - see \cite{H}.}

For all we know, the exact symmetry of the SM is given by the rank four unbroken Lie algebra (obtained from $\mathfrak{g}_4$ (\ref{eq234}) by the substitution $B-L\rightarrow Y$):
\be 
\label{eq316}
\mathfrak{a}_4 = su(3)_c \oplus u(1)_Y \oplus u(1)_Q, \, Q=\frac{1}{2}Y + I_3^L = \frac{1}{3}\sum_{j=1}^3 p'_j - \pi'_2.
\ee
The states of the fundamental (anti)fermions are given by the primitive idempotents of $C\ell_{10}$, represented by the $2^5=32$ different products of the five pairs of basic projectors $\pi_{\alpha}^{(')} , p_j^{(')}$ (\ref{eq38}) (\ref{eq229}). The 16 particles can be labeled by the eigenvalues of the pair of superselected charges $(Q, Y )$:

\begin{eqnarray}
\label{eq317}
(\nu_R) = \ell \, \pi_1 \pi_2 = ( 0, 0 ) = \vert \nu_R \rangle \langle \nu_R\vert,  \, \, (\nu_L) = \ell \, \pi'_1 \pi_2 = ( 0, -1 ) = \vert \nu_L \rangle \langle \nu_L \vert, \, \, \nonumber \\
(e_L) =\ell \, \pi_1 \pi'_2 = ( -1,-1 ) = \vert e_L \rangle \langle e_L \vert, \, \, (e_R) =\ell \, \pi'_1 \pi'_2 =  ( -1,-2 )  = \vert e_R \rangle \langle e_R \vert;  \nonumber \\
\ell = (\nu_L) + (e_L) + (\nu_R) + (e_R) = p_1 \, p_2 \, p_3 , \ \ell^2 = \ell , \ {\rm tr} \, \ell = 4. \, \, \, \, \,
\end{eqnarray}

\begin{eqnarray}
\label{eq318}
(u_L^j) = q_j \, \pi'_1 \pi_2 = (\textstyle \frac23,\frac13 ) = \vert u_L^j \rangle \langle u_L^j \vert, \, 
(d_L^j) = q_j \, \pi_1 \pi'_2 = (\textstyle -\frac13,\frac13 ) = \vert d_L^j \rangle \langle d_L^j \vert, \, \,  \nonumber \\
(u_R^j) = q_j \, \pi_1 \pi_2 = (\textstyle \frac23,\frac43 ) = \vert u_R^j \rangle \langle u_R^j \vert, \,
(d_R^j) = q_j \, \pi'_1 \pi'_2 = (\textstyle -\frac13,-\frac23) = \vert d_R^j \rangle \langle d_R^j \vert;  \nonumber \\
q_j = (u_L^j) + (d_L^j) + (u_R^j) + (d_R^j) = p_j \, p'_k \, p'_{\ell} , \ q_i \, q_j = \delta_{ij} q_j , \ {\rm tr} \, q_j = 4 \, \, \, \, \, \,
\end{eqnarray}
$(j,k,\ell) \in {\rm Perm} (1,2,3) , \ q= q_1 + q_2 + q_3 = q^2 , \ {\rm tr} \, q = 12$. (As the colour is unobservable we do not bother to assign to it eigenvalues of the diagonal operators $i\gamma_{13} , i\gamma_{26} , i\gamma_{45}$ that would replace the index j.)  Note that chirality in the particle subspace, $\mathcal{P}\chi=\chi\mathcal{P}$ is determined by the hypercharge:
\be 
\label{319}
\mathcal{P}\chi = \mathcal{P}(\Pi_R - \Pi_L) = \mathcal{P}(-1)^{3Y} .
\ee
The charges $(Q, Y)$ for the corresponding antiparticles have opposite sign. The spectrum of $Y$ and of $2I_3^L = 2Q-Y$ together with the analysis of \cite{BH} allow to complete the group theoretic version of Proposition 3.1.

\smallskip

\noindent {\bf Remark.} -- The factorization of the primitive idempotents (\ref{eq317}) (\ref{eq318}) into bra and kets involves choices. We demand, following \cite{T21}, that they are hermitian conjugate elements of $C\ell_{10}$, homogeneous in $a_{\alpha}^{(*)}$ and $b_j^{(*)}$ such that the kets corresponding to a left(right)chiral {\it particle} contains an odd (respectively even) number of factors. Choosing then $\vert \nu_R \rangle =a_1a_2\ell, \vert \nu_L \rangle = a_1^*\vert \nu_R \rangle$, we find:
\begin{eqnarray}
\label{eq320}
\langle \nu_R \vert &= &\ell a_2^* a_1^* \, \Rightarrow (\nu_R) = \pi_1\pi_2 \ell,  \,  \vert \nu_L \rangle = \pi'_1 a_2  \ell \, , \nonumber \\
\vert e_L \rangle &= &I_-^L \vert \nu_L \rangle = - a_1 \pi'_2 \ell \, , \ \vert e_R \rangle = -a_1^* \vert e_L \rangle = \pi'_1 \pi'_2  \ell = I_-^R \vert \nu_R \rangle ; \nonumber \\
\vert d_L^j \rangle &= &\pi_1 a_2^* \, q_j \, , \ \vert u_L^j \rangle = I_+^L \vert d_L^j \rangle = a_1^* \pi_2 \, q_j \, , \nonumber \\
\vert d_R^j \rangle &= &a_1^* \vert d_L^j \rangle = a_1^* a_2^* \, q_j \, , \ u_R^j = a_1 \vert u_L^j \rangle = \pi_1 \pi_2 \, q_j \, ,
\end{eqnarray}
$q_j = p_j \, p'_k \, p'_{\ell}$, $j,k,\ell \in {\rm Perm} (1,2,3)$, i.e. $q_1 = p_1 \, p'_2 \, p'_3 = p_1 \, p'_3 \, p'_2$ etc. We note that all above kets as well as all primitive idempotents (\ref{eq317}) (\ref{eq318}) obey a system of 5 equations (specific for each particle), $a_{\alpha} \vert \nu_R \rangle = 0 = b_j \vert \nu_R \rangle$, $a_1^* \vert \nu_L \rangle = a_2 \vert \nu_L \rangle = 0 = b_j \vert \nu_L \rangle$, $\alpha = 1,2$, $j = 1,2,3$, etc. so that they are minimal right ideals in accord with the philosophy of Furey \cite{F16}.

\smallskip

The fact that $\nu_R, \bar{\nu}_L$ are not distinguished by the superselected charges has a physical implication: one can consider their coherent superposition as in the now popular theory of a (hypothetical) Majorana neutrino. This suggests the introduction of a \textit{restricted} 15-dimensional \textit{particle subspace}, with projector 
\be
\label{eq321}
{\cal P}_r = {\cal P} - (\nu_R) = q + \ell_r \, , \quad \ell_r = \ell (1-\pi_1 \pi_2) \, .
\ee
 
Theories whose field algebra is a tensor product of a Dirac spinor bundle on a spacetime manifold with a finite dimensional internal space usually encounter the problem of fermion doubling \cite{GIS} (still discussed over 20 years later, \cite{BS}). It was proposed in \cite{DT20} as a remedy to consider the algebra ${\cal P} C\ell_{10} {\cal P}$ where ${\cal P}$ is the projector (\ref{eq230}) on the 16 dimensional particle subspace (including the hypothetical right-handed sterile neutrino).  It is important - and will be essential in the treatment of the Higgs field (Sect. 4) - that the operators $a_\alpha^{(*)}$ and $b_j^{(*)}$ behave quite differently under particle projection. While $a_\alpha^{(*)}$ commute with $\mathcal{P}$ so that 
\be  
\label{eq322} 
\mathcal{P}a_\alpha^{(*)}\mathcal{P}=a_\alpha^{(*)}\mathcal{P}=\mathcal{P}a_\alpha^{(*)}, \, [\mathcal{P}a_\alpha^*, \mathcal{P} a_\beta]_+ = \mathcal{P} \delta_{\alpha \beta},
\ee
the 2-sided particle projection of $b_j^{(*)}$ vanishes:
\be 
\label{eq323}
\mathcal{P} b_j \mathcal{P} = 0 = \mathcal{P} b_j^* \mathcal{P} . \,
\ee
Accordingly, while the generators (\ref{eq39}) of the (electroweak) flavour "left-right symmetry" $su(2)_L \oplus su(2)_R$ just get multiplied by $\mathcal{P}$, the particle subspace projections of the $su(3)_c$ generators take a modified form:
$$
\mathcal{P}b_jb_k^*\mathcal{P} = b_jb_k^*p'_\ell =: B_{jk} \, \mbox{for} \, (j, k, \ell)\in Perm(1, 2, 3), \, B_{jj}-B_{kk}:=q_j-q_k; 
$$
\be 
\label{eq324}
T_a = \textstyle\frac12 B_{jk} \lambda_a^{kj} \,  , \ \lambda_a \in {\cal H}_3 ({\mathbb C}) , \ {\rm tr} \, \lambda_a = 0 , \ {\rm tr} \, \lambda_a \, \lambda_b = 2 \delta_{ab} , \ a,b = 1,\cdots , 8 \, ,
\ee
but still obey the same CR. It makes sense to consider the gauge Lie algebra in the lepton and the quark sectors (or the factors $C\ell_4$ and $C\ell_6$ in $C\ell_{10}$) separately just noting that $\mathcal{P}(B-L)=-1$ for leptons and $\mathcal{P}(B-L) =\frac{1}{3}$ for quarks. It is particularly appropriate to treat the lepton sector by itself when using the restricted particle space as there the flavour oscillators $a_\alpha^{(*)}$ are also modified:
\be 
\label{eq325}
\ell_ra_1^{(*)}\ell_r = a_1^{(*)} \pi'_2 =:A_1^{(*)}, \, \ell_ra_2^{(*)}\ell_r =a_2^{(*)}\pi'_1 =: A_2^{(*)}.
\ee
The operators $A_\alpha^{(*)}$ provide a realization of the four odd generators of the smallest simple Lie superalgebra,
$s\ell(2|1)$, whose even part is $su(2)_L\oplus u(1)_Y$. (For a detailed identification with the standard definition of 
$s\ell(2|1)$ see Sect. 3 of \cite{T21}.) The nonvanishing anticommutators of $A_\alpha^{(*)}$ are:
$$
[A_1, A_1^*]_+ = \pi'_2 = -Q, \, [A_2, A_2^*]_+ = \pi'_1 = Q - Y, \, 
$$
\be 
\label{eq326}
[A_1^*, A_2]_+ = a_2a_1^* = -I_+, \,  [A_1, A_2^*]_+ = - I_-; \, [I_+, I_-] = 2I_3 = 2Q- Y
\ee
(where we are omitting the superscript L on $I_a$). We shall apply the odd generators $A_\alpha^{(*)}$ in defining the Higgs part of a superconnection in Sect. 4. The \textit{minimal associative envelope} $\mathcal{A}_\ell$ of $s\ell(2|1)\subset C\ell_4$ is 9 dimensional. It contains on top of $A_\alpha^{(*)}$ and their anticommutators (\ref{eq326}) the projector $A_1^* A_1=A_2^* A_2=\pi'_1\pi'_2\in C\ell_4$. The resulting \textit{internal space algebra} which leaves out the sterile neutrinos is the direct sum
\be 
\label{eq327}
\mathcal{A} = \mathcal{A}_\ell \otimes \ell \oplus C\ell_4\otimes (\mathcal{A}_{\ell q} +\mathcal{A}_{qq}+\mathcal{A}_{q\ell}), \, \, \mathcal{A}_{xy}=xC\ell_6^0y .
\ee
Here $\mathcal{A}_{qq}$ is effectively the 9 dimensional associative envelope of $u(3)\subset C\ell_6^0$ spanned by $B_{jk}$ and $q_i$ (3.24); $\mathcal{A}_{\ell q}$ is the 3-dimensional space spanned by $\ell b_1b_2q_3$ and its cyclic permutations, $\mathcal{A}_{q\ell}$ is hermitian conjugate to it. The sum of three terms multiplying $C\ell_4$ in (3.27) is isomorphic to the 15-dimensional Lie algebra $su(4)$.

 \bigskip
 
\subsection{Complex structure associated with $R_7$: a comment}

Following \cite{K,FH, K21} we shall display and discuss the symmetry subalgebras of $C\ell_n, n=8, 9, 10$, of the complex structure generated by the Clifford pseudoscalar $\omega_6^R$ corresponding to the right action of the octonions,
\be
\label{eq328}
\omega_6^R = \gamma_1^R \cdots \gamma_6^R \ \mbox{for} \ \gamma_{\alpha}^R = \epsilon \otimes R_{\alpha} \quad \alpha = 1,\cdots ,7 \, .
\ee
Written in terms of the colour projectors $p_j$ and $p'_j$ the hermitian pseudoscalar $i \omega_6^R$ assumes the form:
$$
i \omega_6^R = \textstyle\frac12 ({\cal P}' - {\cal P} - 3 (B-L)) = \ell + q' - \ell' - q \, ,
$$
\be
\label{eq329}
\ell' = p'_1p'_2p'_3, q'=\sum_{j=1}^3q'_j, \, q'_1=p'_1p_2p_3, q'_2=p_1p'_2p_3, q'_3=p_1p_2p'_3
\ee
where we have used
\be
\label{eq330}
L = \ell - \ell' \, , \ 3B = q-q' \, .
\ee
While the term ${\cal P}' - {\cal P}$ (\ref{eq230}) commutes with the entire derivation algebra $spin (6) = su(4)$ of $C\ell_6$, the centralizer of $B-L$ in $su(4)$ is $u(3)$ -- see Proposition A2 in Appendix A. It follows that the commutant of $\omega_6^R$ in $so(8)$ is $u(3) \oplus u(1)$ while its centralizer in $so(9)$ is the gauge Lie algebra $\mathfrak{g}_{\rm SM} = su(3) \oplus su(2) \oplus u(1)$ of the SM; finally, in $so(10)$, $\omega_6^R$ is invariant under the left-right symmetric extension of $\mathfrak{g}_{\rm SM}$ (\cite{FH, K21}),
\be
\label{eq331}
\mathfrak{g}_{\rm LR} = su(3)_c \oplus su(2)_L \oplus su(2)_R \oplus u(1)_{B-L} \, .
\ee
Furthermore, as proven in \cite{K}, the subgroup of $Spin (9)$ that leaves $\omega_6^R$ invariant is precisely the gauge group\footnote{The group $G_{SM}$ was earlier obtained in \cite{TD-V} starting with the  Albert algebra $J_3^8$ (\ref{eq17}).} $G_{\rm SM} = S (U(2) \times U(3))$ (\ref{GSM}) of the SM (with the appropriate ${\mathbb Z}_6$ factored out). One is then tempted to assume that $C\ell_9$, the associative envelope of the Jordan algebra $J_2^8 = {\cal H}_2 ({\mathbb O})$, may play the role of the internal algebra of the SM, corresponding to one generation of fundamental fermions, with $Spin(9)$ as a GUT group \cite{TD,DT}. We shall demonstrate that although $G_{\rm SM}$ appears as a subgroup of $Spin(9)$ its representation, obtained by restricting the (unique) spinor IR {\bf 16} of $Spin (9)$ to $S(U(2)\times U(3))$ only involves $SU(2)$ doublets, so it has no room for $(e_R) , (u_R), (d_R)$ (\ref{eq317}) (\ref{eq318}). We shall see how this comes about when restricting the realization (\ref{eq39}) of ${\bf I}^L$ and ${\bf I}^R$ to $Spin(9) \subset C\ell_9$. It is clear from (\ref{eq39}) that only the sum $a_1 + a_1^* = \gamma_9$ (not $a_1$ and $a_1^*$ separately) belongs to $C\ell_9$. So the $su(2)$ subalgebra of $spin (9)$ corresponds to the diagonal embedding $su(2) \hookrightarrow su(2)_L \oplus su(2)_R$:
$$
I_+ = I_+^L + I_+^R = (a_1^* + a_1) \, a_2 = \gamma_9 \, a_2 , \ I_- = I_-^L + I_-^R = a_2^* \gamma_9
$$
\be
\label{eq332}
2I_3 = 2I_3^L + 2I_3^R = [a_2 , a_2^*] = \pi_2 - \pi'_2 \, .
\ee
In other words, the spinorial IR {\bf 16} of $Spin (9)$ is an eigensubspace of the projector $P_1 = (2I_3^L)^2$. It consists of four $SU(2)_L$ particle doublets and of their right chiral antiparticles. More generally, the only simple orthogonal groups with a pair of inequivalent complex conjugate fundamental IRs, are $Spin (4n+2)$ (see, e.g. \cite{CD}, Proposition 5.2, p. 571). They include $Spin (10)$ but not $Spin(9)$. 

\smallskip


\smallskip

 A direct description of the IR ${\bf 16}_L$ of $Spin (10)$ acting on ${\mathbb C} {\mathbb H} \otimes {\mathbb C} {\mathbb O}$ is given in \cite{FH1}. (Here ${\mathbb C} {\mathbb H}$ and ${\mathbb C} {\mathbb O}$ are a short hand for the complexified quaternions and octonions: ${\mathbb C} {\mathbb H} := {\mathbb C} \otimes_{\mathbb R} {\mathbb H}$.) The right action of ${\mathbb C} {\mathbb H}$ on elements of ${\mathbb C} {\mathbb H} \otimes {\mathbb C} {\mathbb O}$ which commutes with the left acting $spin (10)$, is interpreted in \cite{FH1} as Lorentz $(SL (2,{\mathbb C}))$ transformation of (unconstrained) 2-component Weyl spinors.

\smallskip

The left-right symmetric extension $\mathfrak{g}_{\rm LR}$ (\ref{eq330}) of $\mathfrak{g}_{\rm SM}$ has a long history, starting with \cite{MP} and vividly (with an admitted bias) told in \cite{S17}. It has been recently invigurated in \cite{HH, DHH}. The group $G_{\rm LR}$ was derived by Boyle \cite{B} starting with the  group $E_6$ of determinant preserving linear automorphisms of the complexified Albert algebra ${\mathbb C} J_3^8$  and following the procedure of \cite{TD-V}.

\bigskip

\section{Particle subspace and the Higgs field}\label{sec4}
\setcounter{equation}{0}

\subsection{The Higgs as a scalar part of a superconnection}

The space of differential forms $\Lambda^*= \Lambda^0 + \Lambda^1 + \Lambda^2 + ...$ can be viewed as $\mathbb{Z}_2$ graded setting $\Lambda_{ev}=\Lambda^0 + \Lambda^2 +..., \Lambda_{od}=\Lambda^1+\Lambda^3+...$. Let $M=M_0+M_1$ be a $\mathbb{Z}_2$ graded matrix algebra. A superconnection in the sense of Quillen \cite{Q, MQ} is an element of $\Lambda_{ev}\otimes M_1 + \Lambda_{od}\otimes M_0$, the odd part of the tensor product $\Lambda^*\otimes M$. A critical review of the convoluted history of this notion and its physical implications is given in Sect. IV of \cite{T-M} . (One should also mention the neat exposition of \cite{R} - in the context of the Weinberg-Salam model with two Higgs doublets.)

Let $D$ be the Yang-Mills connection 1-form of the SM,
$$
D = dx^{\mu} (\partial_{\mu} + A_{\mu} (x)) \, ,
$$
\be
\label{eq41}
i A_{\mu} = W_{\mu}^+ I_+^L + W_{\mu}^- I_-^L + W_{\mu}^3 I_3^L + \frac N2 YB_{\mu} + G_{\mu}^a \, T_a \, ,
\ee
where $Y , {\bf I}^L$ and $T_a$ are given by (\ref{eq314}), (\ref{eq39}) and (\ref{eq324}), respectively, $G_{\mu}^a$ is the gluon field, ${\bf W}_{\!\mu}$ and $B_ {\mu}$ provide an orthonormal basis of electroweak gauge bosons; the normalization constant $N$ wil be fixed in Eq. (\ref{eq413}) below. Then one defines a superconnection ${\mathbb D}$ in \cite{DT20}  involving the chirality $\chi$ (\ref{312}) by
\be
\label{eq42}
{\mathbb D} = \chi D + \Phi \, , \quad \Phi = \sum_{\alpha} (\phi_{\alpha} \, a_{\alpha}^* - \overline\phi_{\alpha} \, a_{\alpha}) \in \mathcal{P}C\ell_{10}^1 \mathcal{P}= \mathcal{P}C\ell_4^1 \, .
\ee
(We omit, for the time being, the projector ${\cal P}$ in $A_{\mu}$ and $\Phi$.) The last equation follows from (\ref{eq323}): the projection on the particle subspace kills the odd part of $C\ell_6$ thus ensuring that the quarks' colour symmetry remains unbroken. The factor $\chi$ (first introduced in this context in \cite{T-M}) ensures the anticommutativity of $\Phi$ and $\chi {\cal D}$ without changing the Yang-Mills curvature $D^2 = (\chi D)^2$.

\smallskip

The projector ${\cal P}$ (\ref{eq230}) on the 16 dimensional particle subspace that includes the hypothetical right chiral neutrino (and is implicit in (\ref{eq42})) was adopted in \cite{DT20}. By contrast, particles are only distinguished  from antiparticles in \cite{T21} if they have different quantum numbers in the Lie algebra of the SM
\be
\label{eq43}
\mathfrak{g}_{SM} = su(3)_c \oplus su(2)_L \oplus u(1)_Y \, .
\ee
Thus, in \cite{T21} ${\cal P}$ is replaced by the 15-dimensional projector ${\cal P}_r = q + \ell_r$ (\ref{eq321}). We have seen that the projected odd operators $A_\alpha^{(*)} = \ell_r a_{\alpha}^{(*)} \ell_r$ give rise to
a realization of the four odd elements of the 8-dimensional simple Lie superalgebra $s\ell (2\vert 1)$ whose even part is the 4-dimensional Lie algebra $u(2)$ of the Weinberg-Salam model of the electroweak interactions. It is precisely the Lie superalgebra proposed in 1979 independently by Ne'eman and by Fairlie \cite{N,F79} (and denoted by them $su(2\vert 1)$) in their attempt to unify $su(2)_L$ with $u(1)_Y$ (and explain the spectrum of the weak hypercharge). Let us stress that the representation space of $s\ell (2 \vert 1)$ consists of the observed left and right chiral leptons (rather than of bosons and fermions like in the popular speculative theories in which the superpartners are hypothetical). Note in passing that the trace of $Y$ on negative chirality leptons $(\nu_L , e_L)$ is equal to its eigenvalue on the unique positive chirality state $(e_R)$ (equal to $-2$) so that only the supertrace of $Y$ vanishes on the lepton (as well as on the quark) space. This observation is useful in the treatment of anomaly cancellation (cf. \cite{T-M20}).

\smallskip

We shall sketch the main steps in the application of the superconnection (\ref{eq42}) to the bosonic sector of the SM emphasizing specific additional hypotheses used on the way (for detailed calculations see \cite{T21}).

\smallskip

The canonical curvature form
\be
\label{eq44}
{\mathbb D}^2 = D^2 + \chi [D,\Phi] + \Phi^2 , \ [D,\Phi] = dx^{\mu} (\partial_{\mu} \Phi + [A_{\mu} , \Phi])
\ee
satisfies the {\it Bianchi identity}
\be
\label{eq45}
{\mathbb D} {\mathbb D}^2 = {\mathbb D}^2 {\mathbb D} \ (\Rightarrow \chi (d\Phi^2 + [A,\phi^2] + [\Phi , D\Phi]_+) = 0) \, ,
\ee
equivalent to the (super) Jacobi identity of our Lie superalgebra. It is important that the Bianchi identity, needed for the consistency of the theory, still holds if we add to ${\mathbb D}^2$ a constant matrix term with a similar structure. Without such a term the Higgs potential would be a multiple of ${\rm Tr} \, \Phi^4$ and would only have a trivial minimum at $\Phi = 0$ yielding no symmetry breaking. The projected form of $\Phi$ (\ref{eq42}) and hence the admissible constant matrix addition to $\Phi^2$ depends on whether we use the projector ${\cal P}$ (as in \cite{DT20}) or $P_r$ (as in \cite{T21}). In the first case we just replace $a_{\alpha}^{(*)}$ with $a_{\alpha}^{(*)} {\cal P}$. In the second, however, the odd generators for leptons and quarks differ and we set:
\be
\label{eq46}
\Phi = \ell (\phi_1 A_1^* - \overline\phi_1 A_1 + \phi_2 \, A_2^* - \overline\phi_2 \, A_2) + \rho q \sum_{\alpha = 1}^2 (\phi_{\alpha} \, a_{\alpha}^* - \overline \phi_{\alpha} \, a_{\alpha}) \, ,
\ee
where $\rho$ (like $N$ in (\ref{eq41})) is a normalization constant that will be fixed later. Recalling that $\ell$ and $q$ are mutually orthogonal $(\ell q = 0 = q\ell , \ \ell + q = {\cal P})$ we find
$$
\Phi^2 = \ell (\phi_1 \overline\phi_2 \, I_+^L + \overline\phi_1 \phi_2 \, I_-^L - \phi_1 \overline\phi_1 \pi'_2 - \phi_2 \, \overline\phi_2 \pi'_1) 
$$
\be
\label{eq47}
- \rho^2 q(\phi_1 \overline\phi_1 + \phi_2 \, \overline\phi_2) \ (\phi_{\alpha} = \phi_{\alpha} (x)) \, .
\ee
This sugggests defining the SM field strength (the extended curvature form) as
\be
\label{eq48}
{\mathbb F} = i ({\mathbb D}^2 + \widehat m^2) \, , \quad \widehat m^2 = m^2 (\ell (1 - \pi_1 \, \pi_2) + \rho^2 q)
\ee
(while $\widehat m^2 = m^2 {\cal P}$ for the 16 dimensional particle subspace of \cite{DT20}). 

\subsection{Higgs potential and mass formulas}

This yields the bosonic Lagrangian (setting $TrX=\frac{1}{4}trX$  - see \cite{T21})
\be
\label{eq49}
{\cal L} (x) = {\rm Tr} \, \Bigl\{ \textstyle\frac12 F_{\mu\nu} F^{\mu\nu} - (\partial_{\mu} \Phi + [A_{\mu} , \Phi])(\partial^{\mu} \Phi + [A^{\mu} , \Phi])\Bigl\} \, - \, V(\Phi)
\ee
where the Higgs potential $V(\Phi)$ is given by (noting that $Tr\ell_r = \frac{3}{4}$):
\be
\label{eq410}
V(\Phi) = {\rm Tr} \, (\widehat m^2 + \Phi^2)^2 - \textstyle\frac14 m^4 = \textstyle\frac12 (1+6 \rho^4)(\phi \overline \phi - m^2)^2 \, .
\ee
Minimizing $V(\Phi)$ gives the expectation value of the square of $\phi =(\phi_1 , \phi_2)$:
\be
\label{eq411}
\langle \phi \, \overline\phi \rangle = \phi_1^m \, \overline{\phi_1^m} + \phi_2^m \, \overline{\phi_2^m} = m^2 , \ \mbox{for} \ \Phi^m = \sum_{\alpha = 1}^2 \phi^m_\alpha a_\alpha^* (\ell \pi'_{3-\alpha} + \rho q) + c \cdot c \, .
\ee
(The superscript $m$ indicates that $\phi_{\alpha}$ take constant in $x$ values depending on the mass parameter $m$.) The mass spectrum of the gauge bosons is determined by the term $- \, {\rm Tr} \, [A_{\mu} , \Phi][A^{\mu} , \Phi]$ of the Lagrangian (\ref{eq49}) with $A_{\mu}$ and $\Phi$ given by (\ref{eq41}) and (\ref{eq46}) for $\phi_{\alpha} = \phi_{\alpha}^m$. The gluon field $G_{\mu}$ does not contribute to the mass term as $C\ell_6^0$ commutes with $C\ell_4^1$. The resulting quadratic form is, in general, not degenerate, so it does not yield a massless photon. It does so however if we assume that $\Phi^m$ is electrically neutral (i.e. commutes with $Q$ (\ref{eq316})):
\be
\label{eq412}
[\Phi^m , Q] = 0 \Rightarrow \phi_2^m = 0 \ (= \overline{\phi_2^m}) \, .
\ee

The normalization constant $N (= {\rm tg} \, \theta_w)$ is fixed by assuming that $2I_3^L$ and $NY$ are equally normalized:
\be
\label{eq413}
N^2 = \frac{{\rm Tr} \, (2I_3^L)^2}{{\rm Tr} \, Y^2} = \frac35 \left(= ({\rm tg} \, \theta_w)^2 \Leftrightarrow \sin^2 \theta_w = \frac38 \right).
\ee
As $Y(\nu_R) = 0 = I^L_3 (\nu_R)$ this result for the ``Weinberg angle at unification scale'' is independent on whether we use ${\cal P}$ or ${\cal P}_r$. If one takes the trace over the leptonic subspace the result would have been $({\rm tg} \, \theta_w)^2 = \frac13 (\Rightarrow \sin \theta_w = \frac12$, \cite{F79}) closer to the measured low energy value.

\smallskip

Demanding, similarly, that the leptonic contribution to $\Phi^2$ is the same as that for a coloured quark (which gives $\rho=1$ for the projector ${\cal P}$) we find
\be
\label{eq414}
\rho^2 = \frac{{\rm Tr} (\ell (1-\pi_1 \pi_2) \Phi^2)}{{\rm Tr}  \, q_j \, \Phi^2} = \frac{Tr (\pi'_1 \pi'_2 \, \phi \, \overline\phi + \pi'_1 \pi_2 \, \phi_2 \, \overline\phi_2 + \pi_1 \pi'_2 \, \phi_1 \, \overline\phi_1)}{4 \, \phi \, \overline\phi} = \frac12 \, .
\ee
The ratio $\frac{m_H^2}{m_W^2}$, on the other hand is found to be
\be
\label{eq415}
\frac{m_H^2}{m_W^2} = 4 \, \frac{1+6 \rho^4}{1+6 \rho^2} = \left\{\begin{matrix}
\mbox{$4$ for $\rho^2 = 1$ (\cite{N}, \cite{DT20})} \\
\mbox{\!\!\!$\frac52$ for $\rho^2 = \frac12$ (\cite{T21})} \qquad\qquad 
\end{matrix} \right. \, .
\ee
The result of \cite{T21}, much closer to the observed value, can also be written in the  form $m_H^2 = 4 \cos^2 \theta_W \, m_W^2$, where $\theta_W$ is the theoretical Weinberg angle (\ref{eq413}).

\section{Outlook}\label{sec5}
\setcounter{equation}{0}

\subsection{Coming to $C\ell_{10}$}

The search for an appropriate choice of a finite dimensional algebra suited to represent the internal space ${\cal F}$ of the SM is still going on. The road to the choice of $C\ell_{10}$, our first step to the restricted algebra $\mathcal{A}$ (\ref{eq327}), has been convoluted.

\smallskip

In view of the lepton-quark correspondence which is embodied in the splitting (\ref{eq11}) of the normed division algebra ${\mathbb O}$ of the octonions, the choice of Dubois-Violette \cite{D16} of the exceptional Jordan algebra ${\cal F} = {\cal H}_3 ({\mathbb O})$ (\ref{eq17}) looked particularly attractive. We realized \cite{TD,TD-V} that the simpler to work with subalgebra
\be
\label{eq15}
J_2^8 = {\cal H}_2 ({\mathbb O}) \subset  {\cal H}_3 ({\mathbb O}) = J_3^8
\ee
corresponds to the observables of one generation of fundamental fermions. The associative envelope of $J_2^8$ is $C\ell_9 = {\mathbb R} [16] \oplus {\mathbb R} [16]$ with associated symmetry group $Spin (9)$. It was proven in \cite{TD-V} that the SM gauge group $G_{\rm SM}$ (\ref{GSM}) is the intersection of $Spin (9)$ with the subgroup of the automorphism group $F_4$ of $J_3^8$ that preserves the splitting (\ref{eq11}), that is the group $\frac{SU(3)\times SU(3)}{\mathbb{Z}_3}\subset F_4$.

\smallskip

So we were inclined to identify $Spin (9)$ as a most economic GUT group. As demonstrated in Sect.~3.4, however, the restriction of the spinor IR {\bf 16} of $Spin (9)$ to its subgroup $G_{\rm SM}$ gives room to only half of the fundamental fermions: the $SU(2)_L$ doublets; the right chiral singlets, $e_R , u_R , d_R$, are left out. It was then recognized that the (octonionic) Clifford algebra $C\ell_{10}$ does the job. The particle interpretation of $C\ell_{10}$ is dictated by the choice of a (maximal) set of five commuting operators in the Pati-Salam Lie subalgebra of $so(10)$ that leaves our complex structure invariant. This led us to presenting all chiral leptons and quarks of one generation as mutually orthogonal idempotents (\ref{eq317}) (\ref{eq318}).

\smallskip

Furey \cite{F} arrived (back in 2018) at the tensor product $C\ell_4\hat{\otimes}C\ell_6$ (\ref{eq111}) following the ${\mathbb R} \otimes {\mathbb C} \otimes {\mathbb H} \otimes {\mathbb O}$ road. In fact, Clifford algebras have arisen as an outgrow of Grassmann algebras and the quaternions\footnote{The Dublin Professor of Astronomy William Rowan Hamilton (1805-1865) and the Stettin Gymnasium teacher Hermann G\"unter Grassmann (1809-1877) published their papers, on quaternions and on ``extensive algebras'', respectively, in the same year 1844. William Kingdom Clifford (1845-1879) combined the two in a ``geometric algebra'' in 1878, a year before his death, aged 33, referring to both of them.}. The 32 products $e_a \varepsilon_{\nu} (= \varepsilon_{\nu} e_a)$, $a = 1 , \cdots , 8$ ($e_8 = \un$), $\nu = 0,1,2,3$ of octonion and quaternion units may serve as components of a $Spin (10)$ Dirac (bi)spinor, acted upon by $C\ell_{10}$ (with generators (\ref{eq224}) involving the operators $L_{\alpha}$) -- cf. \cite{FH1}.

\subsection{Two ways to avoid fermion doubling}

There are two inequivalent possibilities to avoid fermion doubling within $C\ell_{10}$. One, adopted in \cite{DT20,T21} and in Sect.~3.3 of the present survey consists in projecting on the particle subspace, which incorporates four $SU(2)_L$ doublets and eight $SU(2)_L$ (right chiral) singlets, with projector (\ref{eq230})
\be
\label{eq53}
{\cal P} = \ell + q = \frac{1-i\omega_6}2 \, , \ \ell = p_1 \, p_2 \, p_3 \, , \ q = q_1 + q_2 + q_3
\ee
(see (\ref{eq322}), (\ref{eq323}) and (\ref{eq324})). Here $\omega_6$ is the $C\ell_6$ pseudoscalar, the distinguished complex structure, used in \cite{FH} as a first step in the ``cascade of symmetry breakings''. The particle projector (\ref{eq53}) is only invariant under the Pati-Salam subgroup (\ref{eq18}) of $Spin (10)$. The more popular alternative, adopted in \cite{FH1}, projects on left chiral fermions (4 particle doublets and 8 antiparticle singlets) with projector $\Pi_L$, defined in terms of the $C\ell_{10}$ chirality $\chi = i\omega_{10}$:
\be
\label{eq54}
\Pi_L = \frac{1-\chi}2 ={\cal P} P_1 + {\cal P}' P'_1 \quad ({\cal P} + {\cal P}' = 1 = P_1 + P'_1) \, ,
\ee
invariant under the entire $Spin (10)$; here $P_1$ projects on $SU(2)_L$ doublets,  (cf. (\ref{eq310})). The components of the resulting ${\bf 16}_L$ are viewed in \cite{FH1} as Weyl spinors; the right action of (complexified) quaternions (which commutes with the left $spin (10)$ action) is interpreted as an $s\ell (2,{\mathbb C})$ (Lorentz) transformation.

\smallskip

The difference of the two approaches which can be labeled by the projectors ${\cal P}$ and $\Pi_L$ (on left and right particles and on left particles and antiparticles, respectively) has implications in the treatment of the generalized connection (including the Higgs) and anomalies. Thus, for the $\Pi_L$ (anti)leptons $(\nu_L , e_L)$, $\overline e_L$, $\overline{\nu}_L$ we have vanishing trace of the hypercharge, ${\rm tr} \, \Pi_L Y = 0$. For ${\cal P}$ leptons,  $(\nu_L , e_L), \nu_R , e_R$, the traces of the left and right chiral hypercharge are equal: ${\rm tr} ({\cal P} \Pi_L Y) = -2 = {\rm tr} ({\cal P} \Pi_R Y)$, so that, as noted in Sect.~4.1, only the supertrace vanishes in this case. The associated Lie superalgebra fits ideally Quillen's notion of super connection. A real ``physical difference'' only appears under the assumption that the electroweak hypercharge is superselected and $\mathcal{P}$ is replaced by the restricted projector ${\cal P}_r$ on the 15-dimensional particle subspace (with the sterile neutrino $\nu_R$, with vanishing hypercharge, excluded). Then the leptonic (electroweak) part of the SM is governed by the Lie superalgebra $s\ell (2 \vert 1)$, whose four odd generators are given by third degree monomials in $a_{\alpha}^{(*)}$, the $C\ell_4$ Fermi oscillators. The replacement of $\ell$ by $\ell_r$ breaks the quark-lepton symmetry: while each coloured quark $q_j$ appears in four flavours, the colourless leptons are just three. This yields a relative normalization factor between the quark and leptonic projection of the Higgs field and allows to derive (in \cite{T21}) the relation 
\be
\label{eq55}
m_H^2 = \textstyle \frac52 \, m_W^2 = 4 \cos^2 \theta_{\rm th} \, m_W^2 \, ,
\ee
where $\theta_{\rm th}$ is the {\it theoretical} Weinberg angle, such that ${\rm tg}^2 \, \theta_W = \frac35$. The relation (\ref{eq55}) is satisfied within $1\%$ accuracy by the observed Higgs and $W^{\pm}$ masses.

\subsection{Summary and discussion; a challenge}

 After the pioneering work of Feza G\"ursey and collaborators of the 1970's, Geoffrey Dixon devoted to division algebras over 30 years, followed by Cohl Furey since the 2010's. The Clifford algebra approach to unification, coupled to fermionic creation and annihilation operators, has also been pursued, since the late 1970's, by the Italian group around Roberto Caslbuoni. The notion of superconnection was anticipated and applied to the Weinberg-Salam model during the first decade of the creation of the SM, as well.  Thus the basic ingredients of our endeavour have been with us for some 50 years. The pretended new features of the present survey concern certain details.  Here belong:

- The interpretation of the Clifford pseudoscalar $\omega_6$ as $i(\mathcal{P}-\mathcal{P}')$ where $\mathcal{P}$ and 
$\mathcal{P}'$ are the particle and antiparticle projection operators.

- The realization that the projected Clifford algebra 
\be 
\mathcal{P} C\ell_{10}\mathcal{P} = C\ell_4\otimes C\ell_6^0,
\ee
only involves the even part $C\ell_6^0$ of $C\ell_6$, coupled with  the assignment of the Higgs field to the odd part, $C\ell_4^1$, of the first factor, explains the symmetry breaking of the (electroweak) flavour symmetry, while preserving the colour gauge group.

- Exhibiting the role of the sterile neutrino (of the first generation of fundamental fermions) as the vacuum state of the theory. The gauge group of the SM is identified as the maximal subgoup of the Pati-Salam group $G_{PS}$ (1.8) that leaves $\nu_R$ invariant.

- Singling out the \textit{reduced} 15-dimensional \textit{particle subspace} yields a relation between the Higgs and the $W$ boson masses and the theoretical Weinberg angle satisfied within one percent accuracy.

What is missing for completing the ``Algebraic Design of Physics'' -- to quote from the title of the 1994 book by Geoffrey Dixon -- is a true understanding of the {\it three generations} of fundamental fermions. None of the attempts in this direction \cite{F14,D16,T,B, MDW, F23} has brought a clear success. The exceptional Jordan algebra $J_3^8 = {\cal H}_3 ({\mathbb O})$ (\ref{eq17}) with its built in triality was first proposed to this end in \cite{D16} (continued in \cite{DT}); in its straightforward interpretaton, however, it corresponds to the triple coupling of left and right chiral spinors with a vector in internal space, rather than to three generations of fermions. As recalled in (Sect.~5.2 of) \cite{T} any finite-dimensional unital module over ${\cal H}_3 ({\mathbb O})$ has the (disappointingly unimaginative) form of a tensor product of ${\cal H}_3 ({\mathbb O})$ with a finite dimensional real vector space $E$. It was further suggested there that the dimension of $E$ should be divisible by 3 but the idea was not pursued any further. Boyle \cite{B} proposed to consider the complexified exceptional Jordan algebra whose  group of determinant preserving linear automorpghisms is the compact form of $E_6$. This led to a promising left-right symmetric extension of the gauge group of the SM but the discussion has not yet shed new light on the 3 generation problem. Yet another development, based on the study of indecomposable representations of Lie superalgebras, can be traced back from \cite{TJGG} where only the mathematical machinery has been discussed so far.

\bigskip

\noindent \footnotesize{{\bf Acknowledgments} }

\smallskip

\noindent \footnotesize{Discussions with John Baez, Latham Boyle, Michel Dubois-Violette, Cohl Furey and Kirill Krasnov are gratefully acknowledged. It is a pleasure to thank C\'ecile Gourgues for her expert typing. The author thanks IHES for hospitality and support during the course of this work.The present exposition differs from the one in IHES/P/22/01; it ows much to critical remarks on the earlier version by Cohl Furey and especially by Kirill Krasnov.  The author thanks the anonymous referees for their careful reading of the manuscript and constructive critical remarks which helped improve the presentation in the final version of the paper. }

\vglue1cm

\normalsize
\appendix
\section{Appendix}

\subsection*{Notation for Clifford algebras}

{$C\ell(s, t)$} stands for the Clifford algebra generated by $\gamma_\alpha$ satisfying
$$
[\gamma_\alpha, \gamma_\beta]_+ = 2\eta_{\alpha \beta}, \, \eta_{\alpha \alpha} = 1 \, \mbox{for} \, \alpha=1, ..., s,
\eta_{\alpha \alpha} = -1 \, \mbox{for} \, \alpha =s+1, ..., s+t 
$$ 
$(\eta_{\alpha \beta} = 0$ for $\alpha\neq \beta$). Its automorphism group is the (non-compact for $st\neq 0$) orthogonal group $O(s, t)=O(t, s)$. As internal symmetries correspond to compact gauge groups, we are mainly working with (positive or negative) definite forms and use the abbreviated notation $C\ell_s = C\ell(s, 0)$ and $C\ell_{-t}=C\ell(0, t)$ for the associated Clifford algebras. The even subalgebra $C\ell^0(s, t)$ is defined as the (closed under multiplication) span of products of an even number of $\gamma$ matrices; The odd subspace $C\ell^1(s, t)$ is defined as the (real) span of products of odd numbers of $\gamma$'s
(which is not closed under multiplication).

\subsection*{Inter relations between the $L$, $E$, and $R$ bases of $so(8)$}

The imaginary octonion units $e_1 , \cdots , e_7$ obey the anticommutation relations of $C\ell_{-7}$,
$$
[e_{\alpha} , e_{\beta}]_+ := e_{\alpha} e_{\beta} + e_{\beta} e_{\alpha} = - 2 \, \delta_{\alpha \beta} \, , \ \alpha , \beta = 1,\cdots , 7 \eqno ({\rm A}.1)
$$
and give rise to the seven generators $L_{\alpha} = L_{e_{\alpha}}$ of the Lie algebra $so(8)$:
$$
L_{\alpha 8} := \textstyle \frac12 \, L_{\alpha} =: - L_{8\alpha} \, , \ L_{\alpha \beta} := [L_{\alpha 8} , L_{8\beta}] \in so(7) \subset so(8) \, . \eqno ({\rm A}.2)
$$
For $\alpha \ne \beta$ there is a unique $\gamma$ such that
$$
L_{\alpha} \, e_{\beta} = f_{\alpha\beta\gamma} \, e_{\gamma} = \pm e_{\gamma} \, , \ f_{\alpha \beta \gamma} = - f_{\beta\alpha \gamma} = f_{\gamma \alpha \beta} \, . \eqno ({\rm A}.3)
$$
The {\it structure constants} $f_{\alpha\beta\gamma}$ (\ref{eq29}) (which only take values $0,\pm 1$) obey for different triples $(\alpha , \beta , \gamma)$ the relations
$$
f_{\alpha \beta \gamma} = f_{\alpha + 1 \, \beta + 1 \, \gamma + 2} = f_{2\alpha , 2\beta , 2\gamma} \quad ({\rm mod} \, 7) \, . \eqno ({\rm A}.4)
$$
The list (\ref{eq29}) follows from $f_{124} = 1$ and the first equation (A.4), taking into account relations like $f_{679} \equiv f_{672}$ $({\rm mod} \, 7)$ etc. Note that for $f_{\alpha \beta \gamma} \ne 0 \ \ f_{\alpha \beta \gamma}$ are the structure constants of a (quaternionic) $su(2)$ Lie algebra. They are {\it not} structure constants of $so(7) \subset so(8)$.

\smallskip

Define the involutive outer automorphism $\pi$ of the Lie algebra $so(8)$ by its action (\ref{eq213}) on left and right multiplication $L_{\alpha}$ and $R_{\alpha}$ of octonions by imaginary octonions $\alpha = -\alpha^*$:
$$
\pi (L_{\alpha}) = L_{\alpha} + R_{\alpha} =: T_{\alpha} \, , \ \pi (R_{\alpha}) = -R_{\alpha} \Rightarrow \pi (T_{\alpha}) = L_{\alpha} \, . \eqno ({\rm A}.5)
$$
In the basis (A.1) (A.3) of imaginary octonion units $e_{\alpha}$ ($\alpha = 1,\cdots , 7$), setting $e_8 = \un$ and $L_{\alpha 8} = \frac12 \, L_{\alpha}$ (A.2), $R_{\alpha 8} = \frac12 \, R_{\alpha} = -R_{8\alpha}$, we define $E_{ab}$ by the second relation (\ref{eq214})
$$
E_{ab} \, e_c := \delta_{bc} \, e_a - \delta_{ac} \, e_b \, , \ a,b,c = 1 , \cdots , 8 \quad (e_8 = 1) \, . \eqno ({\rm A}.6)
$$

\medskip

\noindent {\bf Proposition A.1} -- Under the above assumptions/definitions we have
$$
\pi (L_{ab}) = E_{ab} \quad (\mbox{for} \ L_{\alpha\beta} := [L_{\alpha 8} , L_{8\beta}] \, , \ L_{\alpha 8} = \textstyle \frac12 \, L_{\alpha} = - L_{8\alpha}) \, . \eqno ({\rm A}.7)
$$

\medskip

\noindent {\bf Proof.} -- From the first equation (A.5) and from (A.1) (A.2) and (A.6) it follows that
$$
E_{\alpha 8} = L_{\alpha 8} + R_{\alpha 8} = \pi (L_{\alpha 8}) \, .\eqno ({\rm A}.8)
$$
The proposition then follows from the relations
$$
L_{\alpha\beta} = [L_{\alpha 8} , L_{8\beta}] \, , \quad E_{\alpha \beta} = [E_{\alpha 8} , E_{8\beta}] \eqno ({\rm A}.9)
$$
and from the assumption that $\pi$ is a Lie algebra homomorphism.

\bigskip

\noindent {\bf Corollary.} -- From (A.7) and the involutive character of $\pi$ it follows that, conversely,
$$
\pi (E_{ab}) = L_{ab} \, . \eqno ({\rm A}.10)
$$
To each $\alpha = 1,\cdots , 7$ there correspond 3 pairs $\beta \gamma$ such that $L_{\beta \gamma}$ and $E_{\beta \gamma}$ commute with $L_{\alpha}$ and among themselves and allow to express $L_{\alpha} = 2L_{\alpha 8}$ in terms of $E_{\alpha 8}$ and the corresponding $E_{\beta \gamma}$:

\begin{eqnarray}
L_1 &= &2 \, L_{18} = E_{18} - E_{24} - E_{37} - E_{56} \, , \nonumber \\
L_2 &= &2 \, L_{28} = E_{28} + E_{14} - E_{35} - E_{67} \, , \nonumber \\
L_3 &= &2 \, L_{38} = E_{38} + E_{17} + E_{25} - E_{46} \, , \nonumber \\
L_4 &= &2 \, L_{48} = E_{48} - E_{12} + E_{36} - E_{57} \, , \nonumber \\
L_5 &= &2 \, L_{58} = E_{58} + E_{16} - E_{23} - E_{47} \, , \nonumber \\
L_6 &= &2 \, L_{68} = E_{68} - E_{15} + E_{27} - E_{34} \, , \nonumber \\
L_7 &= &2 \, L_{78} = E_{78} - E_{13} - E_{26} - E_{45} \, , \ \mbox{or} \ L_{\alpha} = E_{\alpha 8} - \sum_{\beta < \gamma} f_{\alpha \beta \gamma} \, E_{\beta \gamma} \, . \ ({\rm A}.11) \nonumber 
\end{eqnarray}
Recalling that $E_{ab} = \pi (L_{ab})$ (A.8) and the fact that $\pi$ is involutive, so that $\pi (E_{ab}) = L_{ab}$ (A.10) we deduce, in particular,
$$
2 \, E_{78} = L_{78} - L_{13} - L_{26} - L_{45} \, ,
$$
$$
R_7 = 2 \, E_{78} - 2 \, L_{78} = -L_{78} - L_{13} - L_{26} - L_{45} \, , \eqno ({\rm A}.12)
$$
thus reproducing (\ref{eq216}).

\smallskip

We now proceed to displaying the commutant of $i \omega_6$ and $i\omega_6^R$ in $so(7+j)$, $j=1,2,3$.

\bigskip

\noindent {\bf Proposition A.2} -- While the Lie algebra $spin (6) = su(4)$ commutes with $L_7$, the commutant of $R_7$ (A.12) in $su(4) \subset s\ell (4,{\mathbb C})$ is $u(3) (\subset s\ell (4,{\mathbb C}))$ given by
$$
u(3) = \left\{ \sum_{j,k=1}^3 C_{jk} [b_j^* , b_k] \, ; \ C_{jk} \in {\mathbb C} \, , \ C_{kj} = - \overline{C_{jk}} \right\}
\eqno ({\rm A}.13)
$$
in the fermionic oscillator relalization of $C\ell_6 ({\mathbb C})$ (the bar over $C_{jk}$ standing for complex conjugation).

\bigskip

\noindent {\bf Proof.} -- The fact that $L_7 = 2 \, L_{78}$ commutes with the generators $L_{\alpha \beta}$ ($\alpha , \beta = 1,\cdots ,6$) of $so(6)$ follows from (\ref{eq28}). To find the commutant of $R_7$ (A.12) it is convenient to use the fermionic realization of the complexification $s\ell (4,{\mathbb C})$ of $su(4)$ which is spanned by the 9 commutators $[b_j^* , b_k]$ in (A.13) and the 6 products
$$
b_j \, b_k = -b_k \, b_j \, , \ b_j^* \, b_k^* = -b_k^* \, b_j^* \, , \ j,k = 1,2,3, \ j \ne k \, .\eqno ({\rm A}.14)
$$
The sum $L_{13} + L_{26} + L_{45}$ in (A.12) is a multiple of $B-L$ (\ref{eq310}), the hermitian generator of the centre of $g\ell (3,{\mathbb C})$,
$$
B-L \left( = \frac i3 (\gamma_{13} + \gamma_{26} + \gamma_{45})\right) = \frac13 \sum_{j=1}^3 [b_j^* , b_j ] \, . \eqno ({\rm A}.15)
$$
The relations
$$
[B-L , b_j^* \, b_k^*] = \textstyle \frac23 \, b_j^* \, b_k^* \, , \ [B-L , b_j \, b_k] = - \textstyle\frac23 \, b_j \, b_k \, ,
$$
$$
\left[ \left[B-L , [b_j^* , b_k ] \right]\right] = 0 \, , \ j,k = 1,2,3, \ j \ne k \, , \eqno ({\rm A}.16)
$$
show that the commutant of $B-L$ (and hence of $R_7$) in $su(4)$ is $u(3)$.

\medskip

\noindent {\bf Corollary.} -- The commutant of $\omega_6^R$ in $so(8)$ is $u(3) \oplus u(1)$; the commutant of $\omega_6^R$ in $spin (9)$ is the gauge Lie algebra of the SM:
$$
{\cal G}_{\rm SM} = \{ a \in spin (9) \, ; \ [a , \omega_6^R] = 0 \} = u(3) \oplus su(2) \, . \eqno ({\rm A}.17)
$$

\end{document}